\documentclass[12pt]{article}
\pdfoutput=1 
\usepackage{ifpdf}
\usepackage{bbm}
\usepackage{pifont}
\usepackage{mathrsfs}
\usepackage{amsfonts}
\usepackage{graphicx}
\usepackage{latexsym}
\usepackage{epsfig}
\usepackage{amsmath}
\usepackage{amssymb}
\usepackage{color}
\usepackage{amsthm}
\usepackage{natbib}

\usepackage{caption}
\usepackage{subcaption}
\usepackage{mathtools}
\usepackage{graphicx}
\usepackage[toc,page]{appendix}
\usepackage{multirow}
\usepackage{marginnote}
\usepackage{changepage}
\usepackage[usenames,dvipsnames]{xcolor}
\usepackage{hyperref}

\newcommand\lars{LARS}

\newcommand\flars{functional LARS}

\newcommand{\var}{\mbox{Var}}
\newcommand{\cov}{\mbox{Cov}}
\newcommand{\cor}{\mbox{Cor}}
\newcommand{\tr}{\mbox{tr}}
\newcommand{\sd}{\mbox{SD}}
\newcommand{\E}{\mbox{$\mathbb{E}$}}

\DeclareMathOperator*{\Max}{max}

\newcommand\mytilde{\tilde}

\newcommand\fls{fLARS}


\begin{document}
\setlength{\baselineskip}{24pt}
\renewcommand{\baselinestretch}{1.2}

\title{Nonlinear Mixed-effects Scalar-on-function Models and Variable Selection for Kinematic Upper Limb Movement Data}

\author{Cheng, Yafeng \\ MRC Biostatistics Unit \\
Shi, Jian Qing \thanks{Correspondence to: Dr J. Q. Shi, School of Mathematics \& Statistics, Newcastle University, UK, j.q.shi@ncl.ac.uk.}\\ School of Mathematics \& Statistics, Newcastle University, UK \\
Eyre, Janet \\ Institute of Neurosciences, Newcastle University, UK
}
\maketitle

\noindent {\bf{Abstract:}} 
This paper arises from collaborative research the aim of which was to model clinical assessments of upper limb function after stroke using 3D kinematic data. We present a new nonlinear mixed-effects scalar-on-function regression model with a Gaussian process prior focusing on variable selection from large number of candidates including both scalar and function variables. A novel variable selection algorithm has been developed, namely  functional least angle regression (fLARS).  As they are essential for this algorithm, we studied the representation of functional variables with different methods and the correlation between a scalar and a group of mixed scalar and functional variables. We also propose two new stopping rules for practical usage.
This algorithm is able to do variable selection when the number of variables is larger than the sample size. It is efficient and accurate for both variable selection and parameter estimation. Our comprehensive simulation study showed that the method is superior to other existing variable selection methods.  When the algorithm was applied to the analysis of the 3D kinetic movement data the use of the non linear random-effects model and the function variables significantly improved the prediction accuracy for the clinical assessment.

\par

\fontsize{10.95}{14pt plus.8pt minus .6pt}\selectfont

\section{Introduction}

Stroke has emerged as a major global health problem in terms of both death and major disability, that will only continue to increase over the next 20 years as the population ages \citep{Donnan:2008, murray2013disability}. 16 million people worldwide currently suffer a first-time stroke each year, more than 12 million survive. Hemiparesis, a detrimental consequence that many stroke survivors face, is the partial paralysis of one side of the body that occurs due to the brain injury. It is remarkably prevalent: acute cases of hemiparesis are present in 80\% of stroke survivors \citep{Party:2012}. Six months after a stroke, 50-70\% of stroke survivors have persisting hemiparesis \citep{CDC:2012, Kelly:2003}. Studies have consistently demonstrated significant, therapy induced improvements in upper limb function can be achieved, but only with intense, repetitive and challenging practice \citep{Langhorne:2009}. Limited resources, specifically lack of therapist time, are the main barriers to implementation of this evidence-base for stroke rehabilitation \citep{Party:2012}. Conventional rehabilitation programs carried out at home, unsupervised by therapists, are fraught with low compliance \citep{Touillet:2010}. Video games increase compliance since the focus is on game play and fun and not on impairment \citep{rhodes2009predicting}.  The aim of the study reported here is to derive clinically relevant measures of upper limb function from analysis of the movements made by patients whilst playing action-video games in their own home. This will facilitate remote delivery and monitoring by therapists of patients carrying out therapy at home using video games.


In our collaborative research, a home based remote monitoring system has been developed (see \cite{Serradilla:2014, Shi:2013}). An assessment game, including 38 movements, has been designed. Patients after stroke play the assessment game at their home without any supervision by therapists. The 3-dimensional position data and 4-dimensional orientation data (in the format of quaternion) for each movement are recorded and transferred to the cloud. The data is then used to estimate the recovery level of upper limbs for patients. The recovery curve for each patient is constructed and assessed by therapists. This enables therapists to monitor patients recovery and adjust therapy accordingly. 

The controllers used in the assessment game record signal data with 60 Hz frequency \citep{Shi:2013, Serradilla:2014}. These signals can be thought of as densely sampled functional variables. For each of the movement, we get 14 functional variables. In addition to the functional variables, we also have patients personal information and seven summary statistics calculated for each of the movement. The total number of variables is over 700. We also need to consider the heterogeneity between different patients and the nonlinearity between the response and the covariates. Modelling is therefore very challenging. We considered the following complex scalar-on-function regression model:
\begin{align}
y_{i,d}=\beta_0+ \mathbf{z}_{i,d}\boldsymbol\gamma+\sum_{j=1}^J\int x_{i,d,j}(t)\beta_j(t) dt+g(\boldsymbol\phi_{i,d})+\epsilon_{i,d} \label{ME1f}\\
g(\boldsymbol\phi_i)\sim GP(0,\kappa(\boldsymbol\phi_{i,d},\boldsymbol\phi_{i,d'}; \boldsymbol\theta))~~~\epsilon_{i,d}\sim N(0,\sigma^2), \nonumber
\end{align}
where $y_{i,d}$ stands for the recovery level of upper limbs for the $d$-th visit for the $i$-th patient. Clinically assessed CAHAI scores are used to measure recovery level \citep{Barreca:2005}; both $\mathbf{z}_{i,d}$ and $\boldsymbol\phi_{i,d}$ are scalar variables, and $x_{i,d,j}(t)$ is a functional variable. Function $g(\boldsymbol\phi_i)$ is an unknown nonlinear model and a nonparametric Bayesian approach with Gaussian process prior is used to fit the model.

There have been a number of studies in the area of functional linear regression with scalar response and functional predictors. \cite{Cardot:1999, Reiss:2007} applied functional principle component analysis with the univariate functional linear regression with scalar response. \cite{muller2005generalized} discussed generalized linear regression with scalar response and univariate functional predictor in detail. \cite{Crainiceanu:2009} proposed generalized multilevel functional regression following the work from the multilevel FPCA \citep{Di:2009}. 

We pose three novel contributions in the proposed model \eqref{ME1f} for scalar-on-function regression problem. First we use Gaussian process regression to model the nonlinear random-effects. Secondly we propose functional LARS (fLARS), a new algorithm, which can be used to select both scalar and functional variables in \eqref{ME1f} from a large number of candidates. Thirdly, we propose a new efficient method to represent functional variables.

Gaussian process regression is a Bayesian nonparametric model using Gaussian process prior for an unknown nonlinear regression model. It has been widely used in machine learning \citep{Rasmussen:2006}, statistics \citep{Shi:2008, Shi:2011, gramacy2012gaussian, wang2014generalized} and others. It can also be used to model nonlinear random-effects \citep{Shi:2012}. We apply this method in this model to capture the nonlinear variation among different patients.

In the area of multivariate functional regression, a number of variable selection methods have been proposed. For example,  \cite{Matsui:2011,Mingotti:2013,Gertheiss:2013,kayano2015gene} proposed variable selection algorithms for functional linear regression models or functional generalized linear regression models based on the group variable selection methods such as group lasso \citep{Yuan:2006}, group SCAD \citep{Wang:2007} and group elastic net \citep{Zou:2005}. \cite{muller2012functional} proposed the functional additive model, which allows multiple functional variables in the model, but did not propose a corresponding variable selection method.  \cite{fan2013functional} proposed a variable selection algorithm for functional additive models by using group lasso. Their algorithm can handle both linear and non-linear problems, and can select a large number of candidate variables. One common feature in the above proposed algorithms is group variable selection using penalized likelihood algorithms. However with such algorithms, the computational cost increases as the number of candidate variables increases and the estimation and selection accuracy of the functional coefficients become less reliable. 

We propose a new variable selection method, namely functional LARS or fLARS, for the complex scalar-on-function regression model \eqref{ME1f}. 
An extension of the least angle regression model (LARS) \citep{Efron:2003}, it enables variable selection from both mixed scalar and functions. The use of specifically designed stopping rules limits the increase in computational time as the number of candidate variables increases, particularly when the number of variables is larger than the sample size.

To further increase computational efficiency, we propose to use Gaussian quadrature to represent the functional variables and coefficients. Finally to maximise the accuracy of parameter estimation we use the roughness penalty and generalised cross validation.

This paper will be organized as following. We describe the new \fls\ algorithm in Section~\ref{section:flars}. It includes different representation methods of the functional objects, the method to calculate the correlation between scalar and functional variables, stopping rules and some results from the simulation study. Parameter estimation and prediction for model \eqref{ME1f} are discussed in Section~\ref{section:inference}. Numerical results of the application to 3D kinematic data for predicting clinical assessments are reported in Section~\ref{NumericalComparisonHICF}. Finally, some concluding remarks are given in Section~\ref{conclusion}.

\section{Functional LARS algorithm}\label{section:flars}
\subsection{Representation of functional objects}\label{integretion}
In functional regression problem such as Eqn~\eqref{ME1f}, the projection of a functional variable is achieved by integration. We use the following equation for illustration:
\begin{align}
u=\int x(t)\beta(t) dt. \label{Eqn:IntIllu}
\end{align}
Here $u$ is a scalar variable; $x(t)$ is a functional variable and $\beta(t)$ is the corresponding functional coefficient. It is impractical to have functional objects stored as infinite dimensional objects for calculation, so this integration is obtained by numerical approximation. We will briefly discuss the difficulties encountered when applying the most commonly used current models of approximation and our proposed novel solution. We also summarise a unified formula by using three of the methods.

The first commonly used approximation is to use some representative data points (RDP). Suppose that the functional variable $x(t)$ has $q$ representative data points. That is $x(t)=\Big(x(t_1), x(t_2), \ldots, x(t_q)\Big)\triangleq \mathbf{X}_{n\times q}.$ The number $q$ is usually large (100 in our simulation study) and $t_1, \ldots t_q$ are evenly distributed. The coefficient $\beta(t)$ should be observed at the same time points as the functional variable. We can denote $\beta(t)$ as a $p$-dimensional vector $\boldsymbol{\beta}$. It was applied for example in \cite{Leurgans:1993}.

To calculate the second order derivative of $\beta(t)$ for Model~\eqref{ME1f}, we used the finite difference, as applied by \cite{Tibshirani:2005} for first order derivative in fused lasso. Thus:
$\mytilde{\beta}''\approx \boldsymbol{\beta}L^T$, where $L$ is defined as:
\[
L=\left(\begin{array}{ccccccc}
1 &  -2 & 1 & 0  & 0 & 0 & \ldots\\
0  & 1 &  -2 & 1 & 0 & 0 & \ldots\\
\ldots & \ldots & \ldots & \ldots & \ldots & \ldots & \ldots 
\end{array}\right)*\frac{1}{\delta t},
\]
and $\delta t$ is the difference of time between the consecutive data points. This $L$ is from the centred differences formula. The value of $\delta t$ is arbitrary, thus we set it as 1 for convenience. The main limitation of this solution is  high computation cost.

The second method is to use basis functions (BF). This is probably the most commonly used method currently. Suppose $\Phi_q(t)$ are known basis functions that are second order differentiable. We can represent the functional objects by a finite set of basis functions. For example, $\beta(t)
\approx\sum_{q=1}^{Q}\mytilde{C}_{\beta,q}\Phi_q(t)$,
where $\tilde{C}_{\beta,q}$ is the coefficient and $Q$ basis functions are used. We can calculate the second order derivative of the basis functions when we need the second order derivative of $\beta(t)$. Usually we use the same basis functions for all functional objects. However, for different functional variables different sets of basis functions may be required, which is expensive in computational time for both inference and implementation.

Other methods, for example functional principle component analysis \citep{Hall:2006} and functional partial least squares \citep{Reiss:2007,febrero2015functional}, are also popular.

Our novel solution is the use of Gaussian quadrature (GQ) \citep{abramowitz1964handbook}.  Since GQ uses a small set of discrete data points it can be thought of as an extension of the RDP method. The advantage of using GQ method is its efficiency compared to the original RDP method. Depending on the number of points used, the calculation will also be faster than that using BF while giving similar
estimation accuracy.

Many different versions of GQ have been developed and shown to produce accurate approximations in one-dimensional, numerical integration.  The basic formula of GQ is $\int_{-1}^1 f(t) dt \approx \sum_{q=1}^Q w_q f(t_q)$, where the bound [-1,1] is specific to some GQ for example, Gauss-Legendre or Chebyshev-Gauss. Other GQ solutions may have different polynomial functions and ranges for integration and therefore corresponding weights and abscissae.  We propose to use Gauss-Legendre in this paper.


By using Gaussian quadrature, the integration~\eqref{Eqn:IntIllu} can be written as:
\begin{align*}
u=\frac{1}{2}*\int_{-1}^1 x(t)\beta(t) dt = \frac{1}{2}*\mathbf{X}W\mytilde{\beta}, 
\end{align*}
where $W$ is a diagonal matrix. The diagonal entries have weights $w_q$'s at the points closest to the abscissas, and 0 everywhere else. The second order derivative of $\beta(t)$ can be obtained in the similar way to the RDP method. The details are in the Appendix~\ref{GQapp}.


We have combined the three methods of approximation, namely RDP. BF, GQ to both the projection of the functional variable and also to the roughness penalty function:
\begin{align*}
\int x(t)\beta(t) dt ~~~\&~~\int [\beta''(t)]^2 dt.
\end{align*}
We will represent all the functional variables $x(t)$ by the RDP method. There are two benefits of doing this. First, it is very easy to achieve. Second, there are no assumptions made for the data set. This would be useful especially when we have a large number of functional variables. Thus the general expressions of the projection and the roughness penalty function are:
\begin{align}
\int x(t)\beta(t) dt = &\mathbf{X}W \mytilde{C}_{\beta}^T, \label{Eqn:projection}\\
\int [\beta''(t)]^2 dt = &\mytilde{C}_{\beta} W_2 \mytilde{C}_{\beta}^T \label{Eqn:penalty}.
\end{align}
Matrices $W$ and $W_2$ are weight matrices and $\mytilde{C}_{\beta}$ is the unknown coefficient to estimate. If we use RDP method for the functional coefficient, $W=K_I$, which is the diagonal matrix with $1/q$ on the diagonal and $W_2=L^TK_IL$; $\mytilde{C}_{\beta}$ is the discrete vector of the functional coefficient. We have discussed $W$ if we use Gaussian quadrature and $W_2=L^TWL$. Also $\mytilde{C}_{\beta}$ is the functional coefficient, but only the values at or near abscissae can be calculated. If we use BF method, $W=\boldsymbol\Phi/q$; $W_2=\Phi''\Phi''^T/q$; $\mytilde{C}_{\beta}$ is the coefficient for $\beta(t)$. The details can be found in \cite{cheng2016functional}.

\subsection{Correlation Between a Scalar Variable and a Group of Mixed Scalar and Functional Variables}\label{cca:scalar.vs.func}
The key aspect in the least angle regression is the correlation between the response variable and the covariates. For only scalar variables, we can use Pearson's correlation coefficient and also give a projection of the covariates with respect to the response variable. In the group least angle regression \citep{Yuan:2006}, the correlation and the corresponding projection is defined by using the orthonormal basis of the groups of variables. To extend LARS to the functional domain, we need to define the correlation between a scalar variable and functional variables.

We will use the idea from functional canonical correlation analysis (fCCA) in this paper. This type of correlation has been studied by several researchers, for example, \cite{Leurgans:1993} used RDP representation with constraints for smoothness on `curve data'; \cite{Ramsay:2005} applied a roughness penalty with BF representation; \cite{He:2003} discusses a few versions of this analysis; \cite{He:2010} combines functional principle component analysis and canonical correlation analysis in functional linear regression with functional response and predictor. We will adapt the approach by adding roughness constraints to the functional coefficient, and combine it with the three variable representative methods discussed in the previous section.

We first look at the correlation between one scalar variable and one functional variable. If we denote the scalar variable as $y$ and functional variable as $x(t)$, we can define the canonical correlation between them as
\begin{equation}
\rho(x(t),y)=
\Max_{\beta(t),\alpha}\frac{\cov\left(\int x(t)\beta(t)dt,\alpha y\right)}{\sqrt{[\var\left(\int x(t)\beta(t)dt\right)+\lambda\int[\beta^{''}(t)]^2dt][\var(\alpha y)]}} , \label{fccaXY}
\end{equation}
which is equivalent to 
\begin{align*}
\Max_{\beta(t),\alpha} ~~~\cov&\left(\int x(t)\beta(t)dt,\alpha y\right)\\
\text{s.t.~~~~} \var\left(\int x(t)\beta(t)dt\right)&+\lambda[\beta^{''}(t)]^2=1,~~~
\var(\alpha y)=1
\end{align*}

The solution of this optimization would rely on the calculation of the integration involved in the objective function and the constraint. By using the general expressions Eqn~\eqref{Eqn:projection} and Eqn~\eqref{Eqn:penalty}, Eqn~\eqref{fccaXY} becomes:
\begin{align}
\rho(x(t),y)
&=\Max_{\mytilde{C}_{\beta},\alpha}\frac{cov(\mathbf{X}W \mytilde{C}_{\beta}^T, \alpha y)}
{\sqrt{[Var(\mathbf{X}W \mytilde{C}_{\beta}^T)+\lambda \mytilde{C}_{\beta} W_2 \mytilde{C}_{\beta}^T]
[Var(yb)]}} \nonumber\\
&=\Max_{\mytilde{C}_{\beta},\alpha}\frac{\mytilde{C}_{\beta} W^T \mathbf{X}^Ty\alpha}
{\sqrt{[\mytilde{C}_{\beta}(W^T \mathbf{X}^T \mathbf{X} W +\lambda W_2)\mytilde{C}_{\beta}^T][\alpha^2 y^Ty]}}. \label{fccaGen}
\end{align}
Consider the conditions mentioned before Eqn~\eqref{fccaXY}, this maximization problem can be rewritten as maximizing
\begin{equation*}
G=\mytilde{C}_{\beta} W^T \mathbf{X}^Ty\alpha - \frac{1}{2}\rho \mytilde{C}_{\beta}(W^T \mathbf{X}^T \mathbf{X} W+\lambda W_2)\mytilde{C}^T_{\beta}- \frac{1}{2}\rho \alpha^2 y^Ty,
\end{equation*}
where $\rho$ is the correlation coefficient. From the constraint $\var(\alpha y)=1$, $\alpha$ can be obtained straight away: $\alpha=1/\sd(y)$. Given $\alpha$ and the tuning parameter $\lambda$, by using the Lagrange multiplier, we have the following close form:

\begin{align}
\text{correlation:~~~}& \rho^2=\frac{V_{X,y}^T P_{X,X}^{-1} V_{X,y}}{V_y} \label{fccaCorGen}\\
\text{coefficients:~~}& \mytilde{C}_{\beta}=\frac{P_{X,X}^{-1} V_{X,y}}{\rho||y||_2},\label{fccaCoefGen}
\end{align}
where $V_{X,y}=y^T\mathbf{X}W$, $P_{X,X}=W^T \mathbf{X}^T\mathbf{X}W+ \lambda W_2$ and $V_y=y^Ty$. Since we can only get the squared correlation from Eqn~\eqref{fccaCorGen}, we assume that all the correlation is positive for convenience.

The value of the tuning parameter greatly affects the outcome of the calculation. Roughly speaking, the correlation decreases as the value of the tuning parameter increases. As stated in \cite{Ramsay:2005}, the result is meaningless when no constraint is taken for the smoothness of the coefficient for fCCA. On the other hand, when the tuning parameter is too large, the correlation would be reduced to almost zero. We used generalized cross validation (GCV) in our calculation to reduce the computation required.

The smoothing parameter can also help to avoid the singularity problem, however, if the dimension of a group of mixed scalar and functional variables is very large, the singularity problem reappears. To overcome this problem, we introduced a second tuning parameter to form a `sparsity-smoothness' penalty, proposed by \cite{Meier:2009}. It was originally designed for group variables selection and combines the inner product of second order derivatives and the inner product of the original function. The penalty function can be defined as the following:
\begin{equation*}
\mbox{Pen}=\lambda_1 \int[\beta''(t)]^2dt + \lambda_2 \int[\beta(t)]^2dt.
\end{equation*}
\cite{Simon:2012} also mentioned this method when ill conditioning happened with the candidate groups for group lasso. This second tuning parameter can be calculated via cross validation in practice.

We now discuss how to calculate correlation between one scalar variable and a group of mixed scalar and functional variables. The idea is still to Eqn~\eqref{fccaCorGen} with matrix $P_{X,X}$ replaced by block matrices. For example, if both $i$-th and $j$-th variables are functional, the $(i,j)$-th block is:
\[
P_{X_i,X_j}=W^T\mathbf{x}_i^T\mathbf{x}_jW+ \delta_{i,j}\mbox{Pen},
\]
where $\delta_{i,j}=\mathbf{I}$ if $i=j$, and $\mathbf{0}$ otherwise; $\mbox{Pen}$ is the penalty function. The matrix $W$ depends on the choice of the discrete method for $\beta(t)$. If there are one functional variable, $x(t)$ and one scalar variable $z$, the penalized covariance matrix $P_{X,X}$ can be written as:
\[
P_{X,X}= \left( \begin{array}{cc}
W^T\mathbf{x}_i^T\mathbf{x}_jW+\delta_{i,j}\mbox{PEN} & W^T\mathbf{x}_i^Tz\\
\mathbf{x}_jWz & z^Tz
\end{array} \right).
\] 
Similarly, $V_{X,y}$ in Eqn~\eqref{fccaCoefGen} also becomes a block matrix. The $i$-th block is $V_{X_i,y}=W\mathbf{x}_i^Ty$. The above method can be extended to calculate the correlation between two groups of mixed scalar and functional variables. The details can be found in \cite{cheng2016functional}.

\subsection{Functional LARS algorithm}\label{Iter}
We extend the idea of the \lars\ to a functional data analysis framework and propose functional \lars\ algorithm in this section. Our target here is the fixed-effects part in Eqn~\eqref{ME1f} with both scalar and functional candidate variables:
\begin{equation}
y=\sum_{j=1}^J\int x_j(t)\beta_j(t)dt+\sum_{m=1}^Mz_m\gamma_m+\epsilon. \label{FSlinear}
\end{equation}
The integration $\int x_j(t)\beta_j(t)dt$ and the matrix multiplication $\mathbf{z}\gamma$ are both projections to the scalar response in a manner similar to most group variable selection methods. Three representation methods, RDP, GQ and BF, will be considered. The intercept in the regression equation is omitted by assuming that both response and covariates are centred.

Let $x_j(t)$ and $z_m$ be a functional and a scalar variable respectively, and $\mathbf{x}(t)$ and $\mathbf{z}$ be the set that contains all the functional and scalar candidate variables respectively. We also define $A$ as the set of indices of the selected variables and $A^c$ as the set for the remaining candidate variables. Suppose that the residuals obtained from the previous iteration are $r^{(k)}$, where $k$ is the index of the current iteration. Note that for the first iteration, $r^{(1)}=y$, where $y$ is the response variable. Denote $\rho^2(\cdot,y)$ as the squared correlation between a set of variables and a scalar variable $y$. 

The algorithm starts with $k=1$, $r^{(1)}=y$, ${\boldsymbol\beta}(t)=\mathbf{0}$ and $\mathbf{\gamma}=\mathbf{0}$. The first selection is based on the correlation between $r^{(1)}$ and ($\mathbf{x}(t)$,$\mathbf{z}$). The variable that has the largest absolute correlation with $r^{(1)}$ is selected. After the first variable is selected, we carry out the following steps:
\begin{enumerate}
\item Define the direction $u^{(k)}$ to move in by projecting the selected variables to the current residual:
\begin{align*}
u^{(k)}=\frac{\sum_j\int x_j(t)\beta_j^{(k)}(t)dt +\mathbf{z}\boldsymbol\gamma }{\mbox{sd}(\sum_j\int x_j(t)\beta_j^{(k)}(t)dt+\mathbf{z}\boldsymbol\gamma)  }.  
\end{align*}
$u^{(k)}$ must have positive correlation with the residual $r^{(k)}$. Unlike that in group \lars, we normalize the direction vector $u^{(k)}$ using its standard deviation. This normalization will be useful in defining our stopping rule later. The direction of the parameters are estimated in this step.

\item For each remaining variable in $A^c$, compute $\alpha_{l}^{(k)}$ using  
\begin{equation}
\cor(u^{(k)}, r^{(k)}-\alpha_{l}u^{(k)})^2 =\rho^2(x_{l}(t),r^{(k)}-\alpha_{l}u^{(k)})^2~~ l\in A^c  \label{flarsEqn1}
\end{equation}
The variable with the smallest positive $\alpha_{l}^{(k)}$ is selected into the regression equation. Denote the index of that variable as ${l^*}$ and add ${l^*}$ into the set $A$. The distance to move in the direction $u^{(k)}$ is $\alpha_{l^*}^{(k)}$. It is also the scale of the parameters estimated in the previous step.

\item The new residual for next iteration is:
\begin{align}
r^{(k+1)}=r^{(k)}-\alpha_{l^*}^{(k)} u^{(k)}. \label{NewRes}
\end{align}
The functional coefficient up to the $K$-th iteration is the sum of all the coefficients calculated up to and including the current iteration. 
\end{enumerate}

The idea and interpretation of the fLARS are almost the same as the original LARS algorithm. The key difference is owing to Eqn~\eqref{flarsEqn1}. Recall that the \lars\ algorithm uses the following equation to find the distance to move for the direction unit vector with respect to the scalar candidate variable $z$: $\cor(r-\alpha u, z)^2=\cor(r-\alpha u, u)^2$. In our cases, we write this equation in two versions. For functional candidates:
\begin{equation}
\cor\left(r-\alpha u, \int x(t)\beta(t)dt\right)^2/N_f=\cor(r-\alpha u, u)^2/N_u, \label{flarsAlphaF}
\end{equation}
where the correlation on the left hand side of the equation is calculated by functional canonical correlation analysis. For scalar candidates, we have:
\begin{equation}
\cor\left(r-\alpha u, z\right)^2/N_z=\cor(r-\alpha p, p)^2/N_u, \label{flarsAlphaS}
\end{equation}
where $N_f$, $N_z$ and $N_u$ are all constants for normalization.

We now unify the estimation of $\alpha$ in Eqns~\ref{flarsAlphaF} and \ref{flarsAlphaS}. For functional candidate variables, if we substitute Eqn~\eqref{fccaCorGen} into left hand side of Eqn~\eqref{flarsAlphaF} and expand the right hand side of Eqn~\eqref{flarsAlphaF}, we can get:
\begin{align*}
(r-\alpha u)^T\bar{S}(r-\alpha u) &= (r-\alpha u)^T\bar{u}(r-\alpha u).
\end{align*}
where $\bar{S}=\frac{\mathbf{x}WK^{-1}W^T\mathbf{x}^T}{N_f}$, $\bar{U}=\frac{u(u^Tu)^{-1}u^T}{N_u}$; $\mathbf{x}$ is the discrete data points of the functional variable $x(t)$. The distance $\alpha$ is estimated by the solution of the following quadratic equation:
\begin{equation}
\alpha^2[u^T(\bar{S}-
\bar{U})u]-2\alpha[r^T(\bar{S}-\bar{U})u]+
[r^T(\bar{S}-\bar{U})r]=0\label{AlphaF}
\end{equation}

For scalar candidate variables, if we expand both sides of Eqn~\eqref{flarsAlphaS}, we have
\begin{align*}
(r-\alpha u)^T\bar{Z}(r-\alpha u) &= (r-\alpha u)^T\bar{U}(r-\alpha u).
\end{align*}
where $\bar{Z}=\frac{z(z^Tz)^{-1}z^T}{N_z}$. The solution of the following quadratic equation is used as the estimation of $\alpha$.
\begin{equation}
\alpha^2[u^T(\bar{Z}-\bar{U})u]-2\alpha[r^T(\bar{X}-\bar{U})u]+
[r^T(\bar{X}-\bar{U})r]=0\label{AlphaS}
\end{equation}
We can normalize the response, scalar and functional covariates when calculating  $\bar{S}$, $\bar{U}$ and $\bar{Z}$, so that the scalar and functional covariates are treated equally. Note that we need to assume that in each iteration the direction of $\beta_j(t)$ for all $j$ stays the same when $\alpha$ changes, in other words, piece-wise linearity, to perform the above equations.

\subsection{Modifications and stopping rules}\label{section:modification}
Here we propose two modifications to make the algorithm more reliable and efficient. The first one can be used when Eqn~\eqref{flarsEqn1} gives no real solution for all candidate variables. The second modification is to remove some variables which become redundant when new variables are added to the selected variable set. 

\subsubsection{Modification I}
In the algorithm, we must calculate $\alpha$'s for all candidate variables in order to perform the selection. However, in some cases, there may be no real solutions for $\alpha$ in \eqref{AlphaF}. This happens when the candidate variables contain very little information about the current residual. More precisely, the correlation between any candidate variable and the current residual is smaller than the correlation between the current direction and the current residual for all possible values of $\alpha$ before $\alpha u$ reaches the OLS solution. Or if $A^c=\varnothing$, there would be no variable to select and hence no distance to calculate for the last iteration.

In both cases, the selection and the calculation of $\alpha$ would fail. Therefore, in order to have a valid $\alpha$ for the current selected variables, we carry out the OLS for the current direction vector. Thus the algorithm is modified by taking the full OLS using the projection of the selected variables in the regression model when $\alpha$ has no real solutions. If there are candidate variables left, the algorithm may still be able to carry on after this step, but the contribution from the new variables would be little.

\subsubsection{Modification II}\label{modification2}
The \lars\ algorithm itself would not reduce the number of selected variables without modification. But, with the suitable modification, it can give lasso solution, which will involve dropping some variables from set $A$. A variable is removed from the regression equation when the sign of its coefficient changes. Changing of the coefficient sign indicates that there is a point at which the coefficient is exactly zero. This implies that the corresponding variable contributes nothing to explaining the variation of the response variable at that point. 

For the functional \lars\ algorithm, the sign change is no longer feasible, since the sign of $\beta(t)$ is difficult to define. An alternative way is to measure the contribution of a variable by calculating the variance of its projection on the response variable, e.g., $\var\left(\int x(t)\beta(t) dt\right)$. Since it is impossible to find the exact point that the variance reaches zero, the variable is removed when the variance is reduced so that it makes little contribution to the total variance. This criterion gives the following two conditions. Firstly, the variance of the projection of the variable is smaller than the maximum variance from the same variable:
\begin{align*}
\var\left(\int x_j(t)\beta_j^{(k)}(t)\right) < \var\left(\int x_j(t)\beta_j^{(k*)}(t)\right)~~\text{for}~~ k*\in 1,\ldots, k-1.
\end{align*}
This is to avoid removing newly selected variables, since the distances to move for the newly selected variables might be small, and leads to small variances of projection. Secondly, the variance of the projection of the variable is less than a certain percentage of the total variance of the response variable:
\begin{align*}
\var\left(\int x_j(t)\beta_j^{(k)}(t)\right) < \kappa\var(y),
\end{align*}
where $\kappa$ is a threshold. In our simulation, 5\% is used. A few thresholds can be tested in the real situation by methods such as $k$-fold cross validation. 

\subsubsection{Stopping rules}
Similar to the \lars\ and group \lars\ algorithms, \flars\ is also able to calculate the full solution path or solution surface more precisely. Practically, the final model should be one of the estimates on the solution path. We can always use leave-one-out cross validation to find the optimal stopping point, but it is very time consuming. Therefore we need to use other stopping rules which are less expensive computationally. Mallow's $C_p$-type criteria \citep{Mallows:1973} have been used in the \lars\ and group \lars. In addition, other traditional methods, including Akaike information criterion (AIC) \citep{Akaike:1998}, Bayesian information criterion (BIC) \cite{Schwarz:1978}, adjusted $R^2$ coefficient can also be used. We will show that these criteria cannot be used in the \flars\ algorithm before we propose a new stopping rule. 

We take $C_p$ type criteria as an example. For a linear regression problem $y=\beta_0+\sum_{j=1}^Jz_j\beta_j+\epsilon$, where $\epsilon\sim N(0,\sigma^2)$. We can assume that all the variables are centred, so that $\beta_0=0$. Thus the response variable $y$ has the distribution $N(\mu, \sigma^2)$, where $\mu=\sum_{j=1}^Jz_j\beta_j$. Suppose that we have $n$ independent sets of observations. 

The estimation of $C_p$ criterion is:
\begin{align}
C_p=\E\Bigg[\frac{(\mathbf{y}-\hat{\boldsymbol\mu})^2}{\sigma^2}\Bigg]-n+2\text{df}, \label{MallowsCp}
\end{align}
where df is the degrees of freedom, defined as:
\begin{align}
\text{df}=\tr\Bigg(\frac{\cov(\hat{\boldsymbol\mu}^T,\mathbf{y}^T)}{\sigma^2}\Bigg)\label{DoF}.
\end{align}

This definition of degrees of freedom traditionally comes from Stein’s unbiased risk estimation (SURE) theory \citep{stein1981estimation}. It is also discussed in \cite{ye:1998} as `generalized degrees of freedom'. \cite{Efron:1997} also mentioned this type of definition. Both the \lars\ and group \lars\ algorithms use Eqn~\eqref{MallowsCp} and Eqn~\eqref{DoF} in their stopping rules. This definition requires bootstraps to calculate the correlation in \eqref{DoF}, and therefore it is computationally expensive. `Hat' matrix can be used to approximate the correlation. We define the degrees of freedom in \flars\ based on the `hat' matrix. Let the `hat' matrix at iteration $k$ be:
\begin{align*}
H_k^*=\frac{H_k \alpha_k}{\text{\sd}(H_k r^{(k)})},
\end{align*}
where $r^{(k)}$ is the residual at the $k$-th iteration and $H_k$ is from Eqn~\eqref{fccaCorGen} and Eqn~\eqref{fccaCoefGen}. More specifically, if $\mathbf{X}$ is the block matrix that combines all selected variables, $H_k=\mathbf{X}W (W^T \mathbf{X}^T\mathbf{X}W+ \lambda_1 W_2 +\lambda_2 W^TW)^{-1} W^T\mathbf{X}^T$. Also we define the `hat' matrix $\bar{H}_K$ after iteration $K$ as
\begin{align}
\bar{H}_K=I-\prod_{k=1}^{K}(I-H_k^*) \label{HatK}.
\end{align}
The degrees of freedom for functional LARS are therefore
\begin{align}
\text{df}^*=\tr(\bar{H}_k). \label{BarHk}
\end{align}
We can use this definition in Eqn~\eqref{MallowsCp} to give a $C_p$ estimate. More details about the calculation of the degrees of freedom is in the Appendix~\ref{dfFLARS}.

The degrees of freedom in \eqref{BarHk} is calculated by using the information from both the response variable and the covariates, and it is representing the relationship between the response variable and the covariates. This is due to the use of the tuning parameters in \fls. Therefore the degrees of freedom do not necessarily increase when the number of variables increases, and vice versa. Thus it cannot provide enough penalty in the conventional information based criteria.

Now we discuss our new stopping rules. The idea is to compare the contributions from each iteration. This contribution can be represented by $\alpha_{l^*}^{(k)} u^{(k)}$ in Eqn~\eqref{NewRes}. Since the direction vectors $u^{(k)}$ are centred unit vectors, the distance $\alpha$ is equivalent to: 
\begin{align*}
\alpha=\alpha||u||_2=||\alpha u||_2= \sd(\alpha u)\sqrt{n-1}. 
\end{align*}

Thus $\alpha$ can represent the variation explained in the current iteration. A very small $\alpha$ means that the current iteration provides very little information. Such a phenomenon can happen in two situations: firstly, the current direction vector $u$ is informative and it is almost equally important, yet very different, as the newly selected variable $Z$ with respect to the residual $r$. Secondly, $u$ is unable to explain much variation in the residual $r$. 

In the latter situation, $u$ would have small correlation with $r$. Therefore it only needs a small distance for $u$ to move to reach the OLS solution with respect to $r$. This indicates that we can stop the algorithm when the distance is very small and correlation is small. Suppose that $Z^{(k+1)}$ is a newly selected variable. We can check the correlation between it and the current residual $r^{(k)}-\alpha^{(k)} u^{(k)}$ to find out how informative $u$ is. If this correlation is small, $Z^{(k+1)}$ would not be useful to explain the variation in the current residual. 
Because
\[
\cor(p^{(k)}, r^{(k)}-\alpha^{(k)} u^{(k)})^2 =\rho(Z^{(k+1)},r^{(k)}-\alpha^{(k)} u^{(k)})^2,
\]
we can use $\cor(u^{(k)}, r^{(k)}-\alpha^{(k)} u^{(k)})^2$ instead of $\rho(Z^{(k+1)},r^{(k)}-\alpha^{(k)} u^{(k)})^2$ to measure the correlation between the newly selected variable and the current residual to avoid the heavy calculation of functional canonical correlation, if $Z^{(k+1)}$ is a functional variable. Let us denote this correlation as $\rho^*$.

\begin{figure}
\centering
\begin{minipage}{.5\textwidth}
  \centering
  \includegraphics[width=0.9\linewidth,height=0.28\textheight]{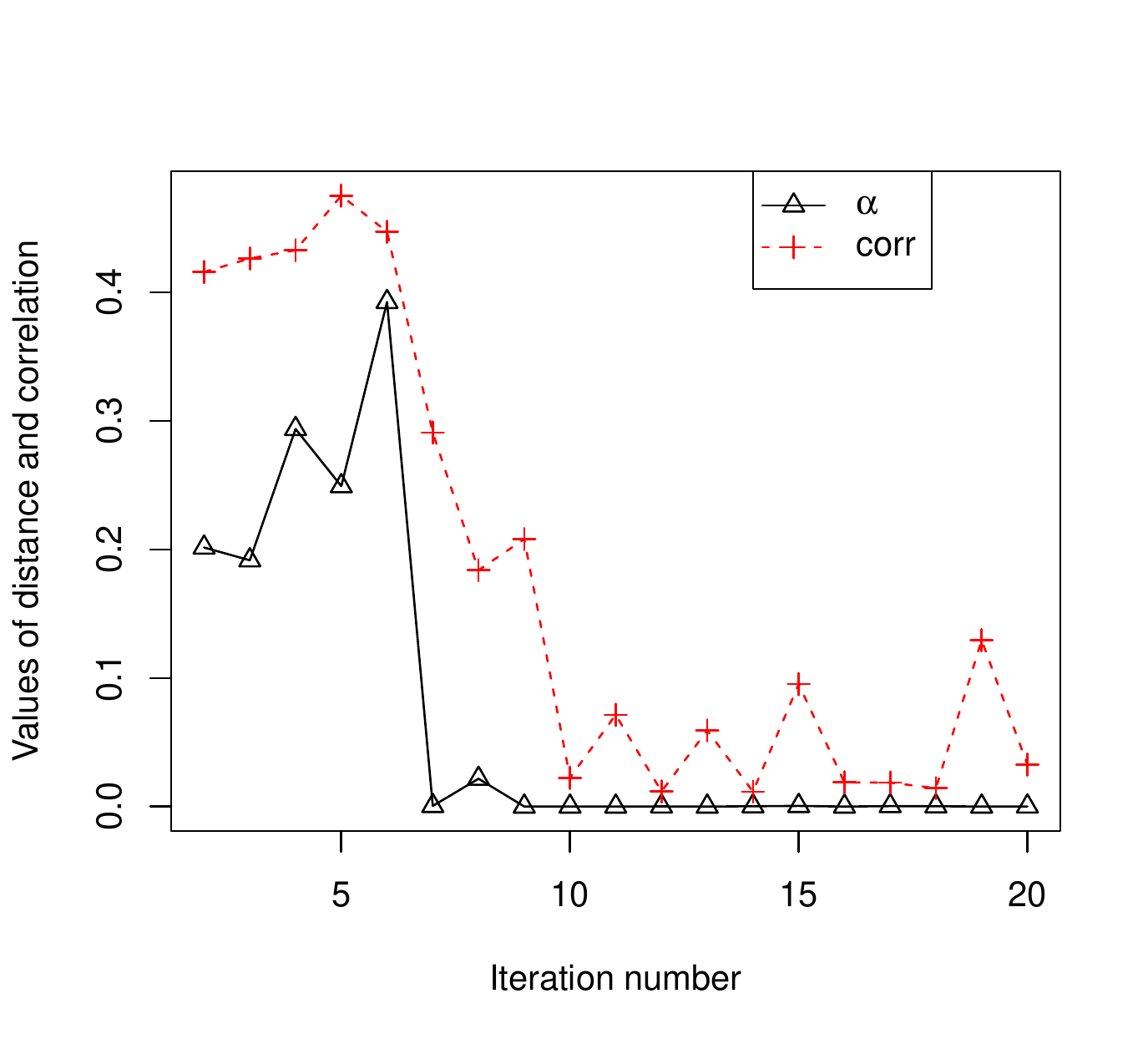}
  \captionof{figure}{Change of $\alpha$ and $\rho^*$.}
  \label{stopping0}
\end{minipage}%
\begin{minipage}{.5\textwidth}
  \centering
  \includegraphics[width=0.9\linewidth,height=0.28\textheight]{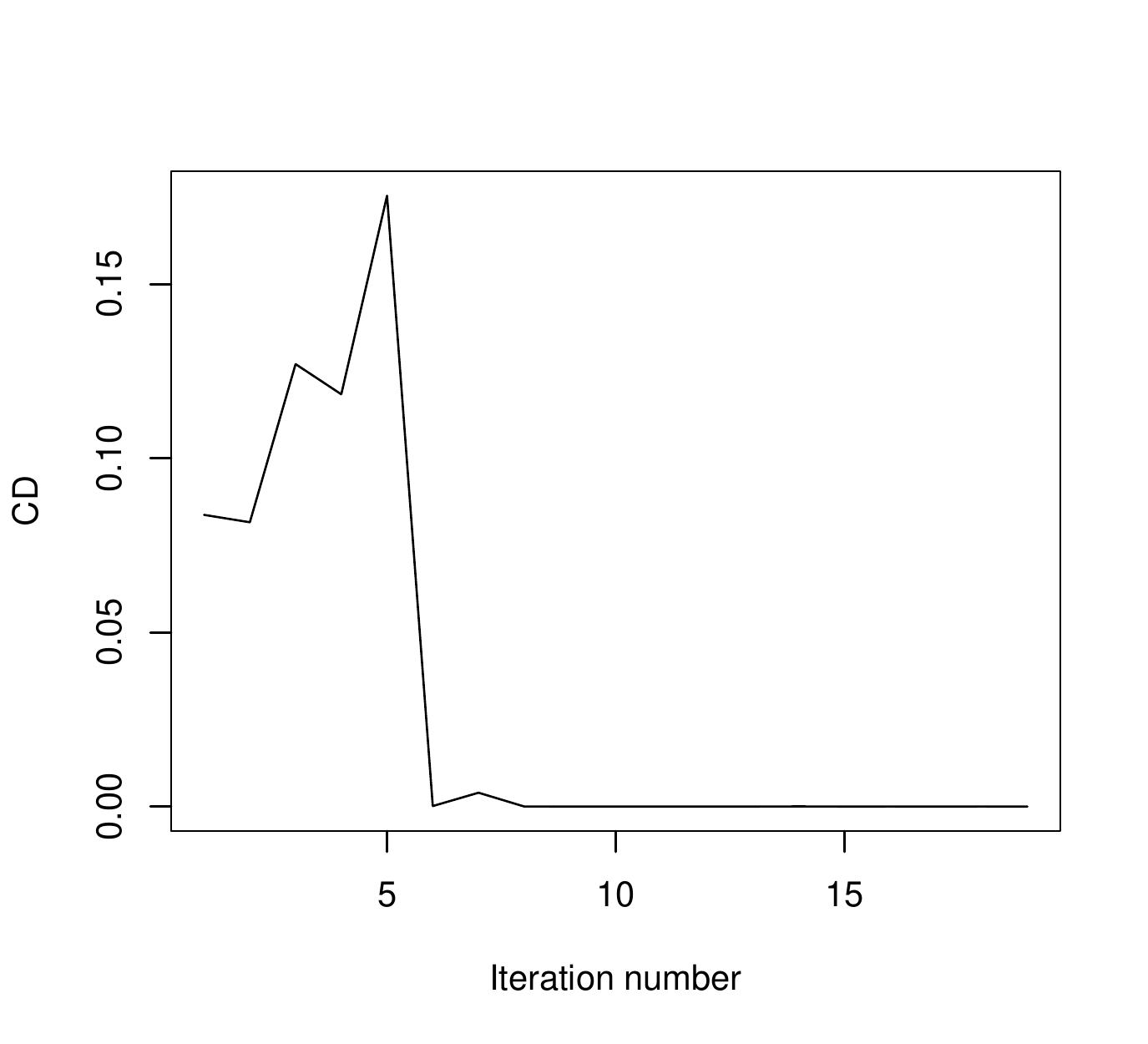}
  \captionof{figure}{Change of `CD'.}
  \label{stopping1}
\end{minipage}
\end{figure}

Thus the algorithm can stop when the newly selected variable has a very small correlation with the current residual and provides a small $\alpha$. For example, Fig~\ref{stopping0} shows the changes of $\alpha$ and the correlations against the iteration number. Note that the iteration number starts from 2, since we have no correlation value for the first iteration. This plot is drawn based on a model with 100 candidate variables and six true variables. Based on the plot, we can see that the distance $\alpha$ reduces to almost 0 after the sixth iteration, and the correlation starts to reduce markedly. The first six selections include all six true variables. Therefore, if we aim to select correctly the variables, we should stop after the sixth iteration. 

We can merge the two requirements to make it easier to use in practice. Let us define `CD', short for correlation times distance, as:
\[
\mbox{CD}_k=\rho_k^* \times  \alpha_k.
\]
We can stop the selection when there appears a clear drop on the `CD' as illustrated in Figure~\ref{stopping1}  after the 6th iteration. Numerically, we can set a threshold, such as 10\% of the maximum `CD'. 

\subsection{Simulation study}
We consider two scenarios in this simulation study: Scenario 1 has seven functional and five scalar candidate variables, while Scenario 2 has 50 functional and 50 scalar candidate variables.

\begin{figure}[!htp]
\centering
\includegraphics[height=1.3in,width=5in]{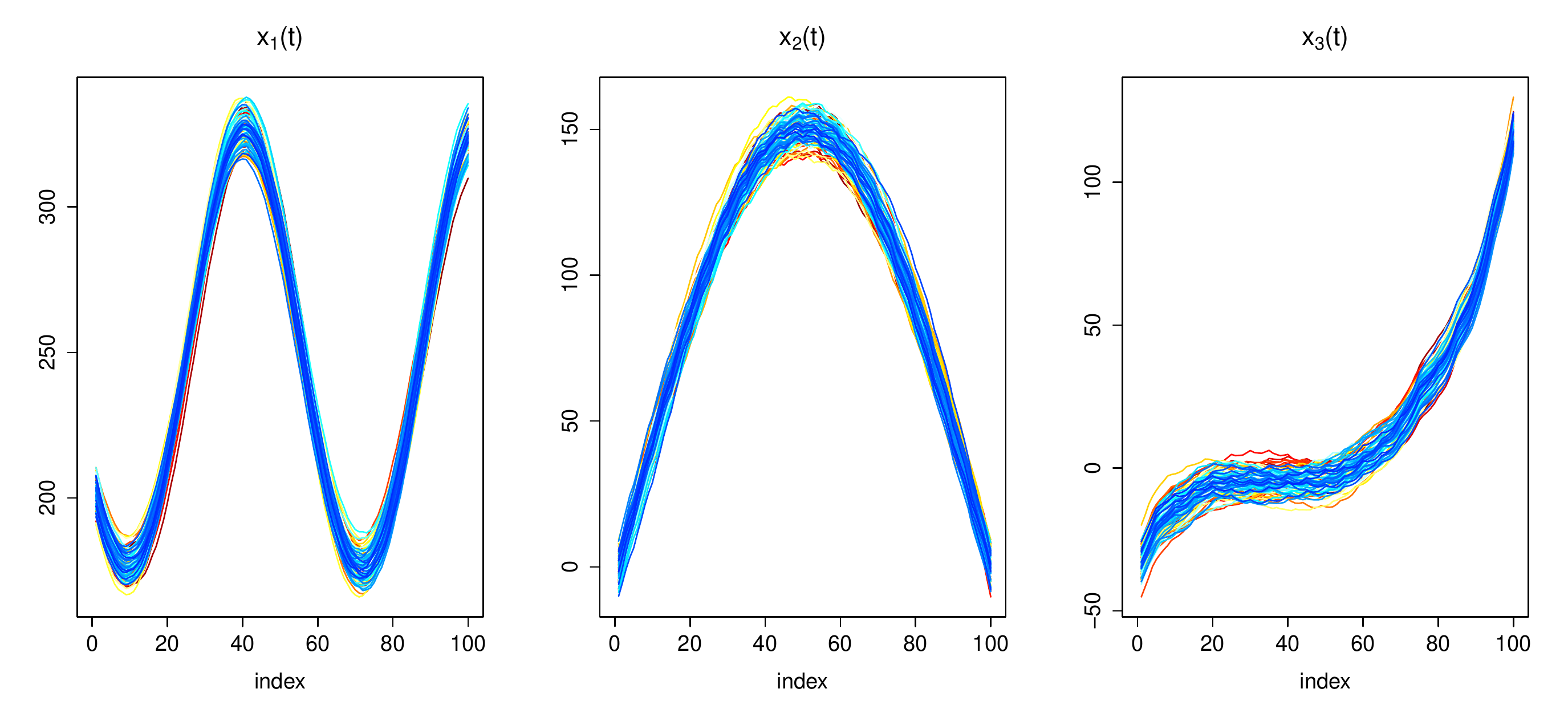}
\includegraphics[height=1.3in,width=5in]{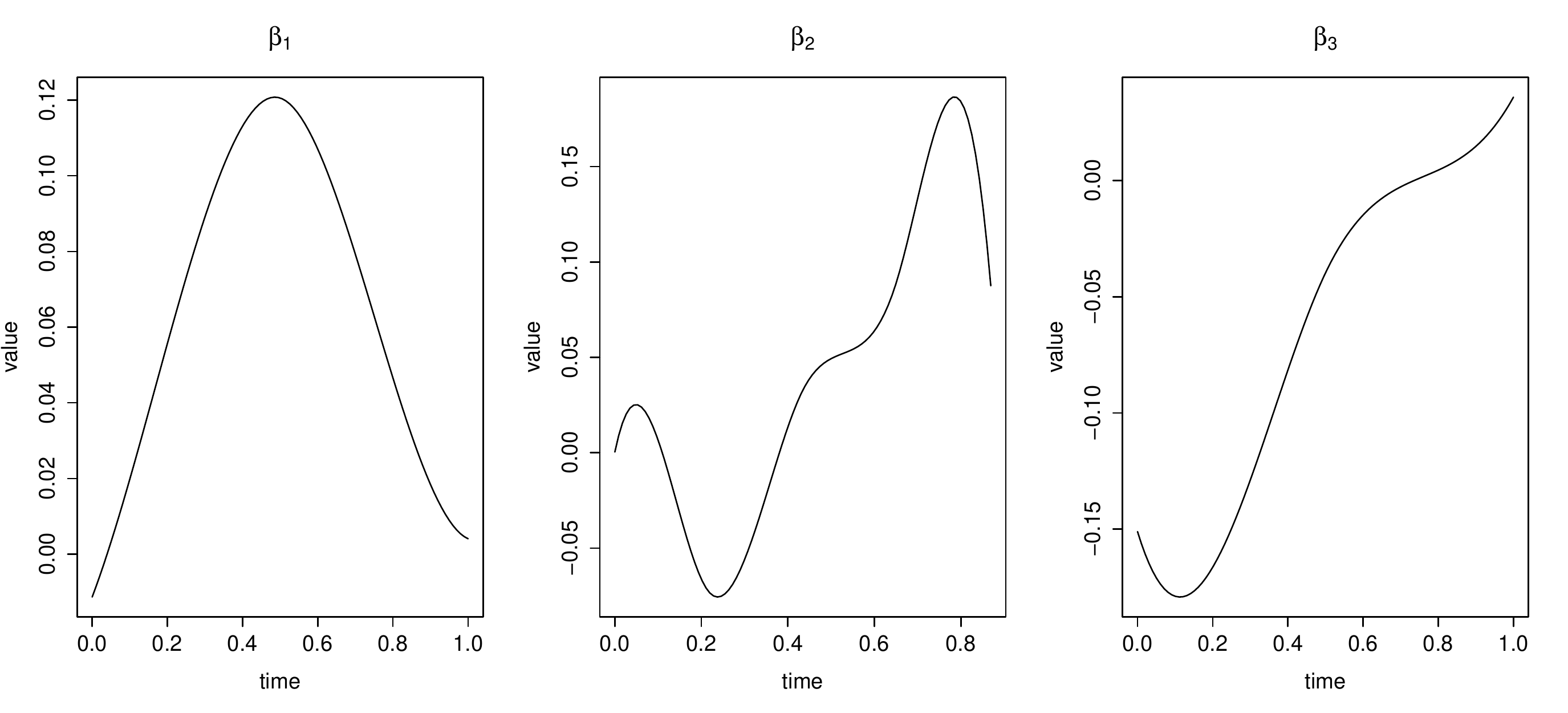}
\caption{The true values of the functional coefficients}
\label{fig:beta_fv_uncor}
\end{figure}

The true model uses three functional variables:
\begin{equation}
y=\mu+\int x_1(t)\beta_1(t) dt +\int x_2(t)\beta_2(t) dt +\int x_3(t)\beta_3(t) dt + z_1\gamma_1 + z_2\gamma_2 + z_3\gamma_3 + \epsilon \label{eqn:TrueFLARS}
\end{equation}
where $\mu$ is the intercept and $\epsilon$ the noise follows a normal distribution with mean 0 and standard deviation 0.05. As an illustrative example, one set of functional variables and the corresponding functional coefficients are shown in Figure~\ref{fig:beta_fv_uncor}. 

We compare two algorithms here with ours. One is using the idea from \cite{Gertheiss:2013} and the other one is using the ideas from \cite{Mingotti:2013}, \cite{Matsui:2011} and \cite{Matsui:2008}. There ideas can be summarised as following: transfer functional variables into groups of variables by basis functions, and apply the regularized group variable selection methods such as group lasso and group SCAD. Because group variable selection algorithms have no requirements on the dimensions of the candidate groups of variables, these algorithms can be easily extended to the case where both scalar variables and functional variables are included in the candidate set. The difference between the two ideas is the way of controlling the smoothness of the functional coefficients. The former ($\mbox{FGL}_P$) uses roughness penalty while the later ($\mbox{FGL}_B$) uses a certain number of basis functions. The roughness penalty gives a more accurate estimation than the later idea, but it requires heavier computation.

For the proposed \fls, three representation methods as discussed in Section~\ref{integretion} are considered. For each of them, three normalization methods are used, i.e., we choose Frobenius norm (Norm), trace (Trace) and identity matrix (Identity) in Eqn~\eqref{AlphaF} and \eqref{AlphaS}. The results for both scenarios are based on 1000 replications.

\subsubsection*{Scenario 1: twelve candidate variables}
Table~\ref{tab:SumThreeAlgoMS2} is a summary of the simulation. The prediction accuracy is shown by root-mean-square-error (RMSE) between the predictions and the simulated values for the test data. The selection accuracy contains two parts: the percentage of true selections or false selections. For example, in this scenario, in one replication, suppose A true variables are selected, together with B other variables, the percentage for true and false selections are $A/(A+B)$ and $B/(A+B)$ respectively. The computation time (average time per replication) is also reported in seconds. For all the algorithms, this is the time when the stopping point is found. For $\mbox{fLARS}$, 100 discrete data points for $RDP$ method, 18 Gaussian quadrature rule for GQ method and 18 basis functions for BF method are used. For $\mbox{FGL}_P$, five fold cross validation is carried out with 21 basis functions for functional variables. For $\mbox{FGL}_B$, we used five fold cross validation and 9 basis functions for functional variables.

The prediction accuracy from different versions of $\mbox{fLARS}$ is generally better than that of $\mbox{FGL}_P$. The prediction result from $\mbox{FGL}_B$ is the worst. Both $\mbox{FGL}_P$ and $\mbox{FGL}_B$ have many more false selections than those from $\mbox{fLARS}$. The fastest algorithm is the $\mbox{fLARS}$ with Gaussian quadrature and identity normalization. The $\mbox{fLARS}$ algorithm with $RDP$ methods are understandably slow due to the high dimension of the representation for each of the functional variables. The $\mbox{FGL}_P$ takes a long time due to the two tuning parameters.

\begin{table}[t]
\centering
\begin{tabular}{|c|c|c|r|r|r|r|}
\hline
\multicolumn{3}{|c|}{Algorithms } & RMSE & True selection (\%) & False selection(\%) & Time (sec) \\
\hline
\multirow{ 9}{*}{$\mbox{fLARS}$} & \multirow{ 3}{*}{$RDP$} & 
		Norm & 0.0586 & 99.74 & 0.26 & 3.1231 \\ 
\cline{3-7}
&  & Trace & 0.0613 & 99.80 & 0.20 & 3.3727 \\ 
\cline{3-7}
&  & Identity & 0.0601 & 99.44 & 0.56 & 3.8276 \\ 
\cline{2-7}
& \multirow{ 3}{*}{$GQ$} & 
 		Norm & 0.0621 & 99.32 & 0.68 & 0.9492 \\ 
\cline{3-7}
&  & Trace & 0.0661 & 98.31 & 1.69 & 0.5288 \\ 
\cline{3-7}
&  & Identity & 0.0630 & 99.03 & 0.97 & 0.5123 \\ 
\cline{2-7}
& \multirow{ 3}{*}{$BF$} & 
  		Norm & 0.0596 & 99.14 & 0.95 & 0.6556 \\ 
\cline{3-7}
&  & Trace & 0.0682 & 95.85 & 4.15 & 1.0814 \\ 
\cline{3-7}
&  & Identity & 0.0594 & 98.90 & 1.10 & 0.9101  \\ 
\hline
\multicolumn{3}{|c|}{$\mbox{FGL}_P$} & 0.0617 & 66.39 & 33.61 & 250.7358 \\ 
\hline
\multicolumn{3}{|c|}{$\mbox{FGL}_B$} & 0.1181 & 80.29 & 19.71 & 1.0692 \\ 
\hline
\end{tabular}
\vspace{0.2cm}
\caption{Comparison of different methods from Scenario 1.}
\label{tab:SumThreeAlgoMS2}
\end{table}

\begin{table}[!htp]
\begin{adjustwidth}{-2.3cm}{}
\begin{tabular}{|c|c|rr|rr|rr|rr|}
  \hline
 & & \multicolumn{2}{c|}{RMSE} & \multicolumn{2}{c}{Expected (\%)} & \multicolumn{2}{|c}{Unexpected (\%)}& \multicolumn{2}{|c|}{Time (sec)} \\ \cline{3-10}
 & & \footnotesize{Unmodified} & \footnotesize{Modified} & \footnotesize{Unmodified} & \footnotesize{Modified}  & \footnotesize{Unmodified} & \footnotesize{Modified} & \footnotesize{Unmodified} & \footnotesize{Modified}\\ \hline
\multirow{ 3}{*}{\shortstack{$\mbox{fLARS}$\\$RDP$}} & Norm & 0.065 & 0.0639 & 99.85 & 99.85 & 0.15 & 0.15 & 14.7855 & 14.3802 \\ 
\cline{2-10}
  & Trace & 0.0877 & 0.0849 & 99.73 & 99.77 & 0.26 & 0.23 & 13.8667 & 13.4114 \\ 
\cline{2-10}
  & Identity & 0.0655 & 0.065 & 99.76 & 99.77 & 0.24 & 0.23 & 15.2081 & 14.7832 \\ 
\hline
\multirow{ 3}{*}{\shortstack{$\mbox{fLARS}$\\$GQ$}} & Norm & 0.0774 & 0.0721 & 98.47 & 99.50 & 0.53 & 0.50 & 3.4819 & 3.6457 \\ 
\cline{2-10}
  & Trace & 0.0926 & 0.0848 & 99.32 & 99.41 & 0.68 & 0.59 & 3.2658 & 3.4915 \\ 
\cline{2-10}
  & Identity & 0.1183 & 0.1128 & 98.51 & 98.67 & 1.49 & 1.33 & 3.4951 & 3.6178 \\ 
\hline
\multirow{ 3}{*}{\shortstack{$\mbox{fLARS}$\\$BF$}} & Norm & 0.0655 & 0.0622 & 99.74 & 99.74 & 0.26 & 0.26 & 5.2638 & 5.5291 \\ 
\cline{2-10}
  & Trace & 0.1002 & 0.0952 & 99.06 & 99.22 & 0.94 & 0.78 & 5.0382 & 5.4151 \\ 
\cline{2-10}
  & Identity & 0.0777 & 0.0745 & 99.44 & 99.46 & 0.56 & 0.54 & 5.0768 & 5.359 \\ 
\hline
\multicolumn{2}{|c|}{$\mbox{FGL}_P$} & \multicolumn{2}{c|}{0.1103} & \multicolumn{2}{c|}{69.66} & \multicolumn{2}{c|}{30.34} & \multicolumn{2}{c|}{1659.4159} \\ 
\hline
\multicolumn{2}{|c|}{$\mbox{FGL}_B$} & \multicolumn{2}{c|}{0.3662} & \multicolumn{2}{c|}{79.57} & \multicolumn{2}{c|}{20.43} & \multicolumn{2}{c|}{9.3553} \\ 
\hline
\end{tabular}
\end{adjustwidth}
\caption{Comparison of the prediction and the selection accuracy for Scenario 2.}
\label{tab:comp7}

\end{table}
\subsubsection*{Scenario 2: one hundred candidate variables}
We show the results in Table~\ref{tab:comp7}. We now include the comparison of the results from unmodified and modified $\mbox{fLARS}$ algorithms. The threshold we used in our simulation for modification II in Section~\ref{modification2} is 0.05.

We can first make similar conclusions as that from the previous scenario about the prediction accuracy, selection accuracy and computational time using different algorithms. In addition, when we have large number of covariates, the modified $\mbox{fLARS}$ is in general better than the unmodified version. The degree of improvement is not great but is consistent across the different data. From our experience by carefully adjusting the threshold even greater accuracy can be obtained, but at the cost of greater computational time.

\section{Inference}\label{section:inference}
Model~\eqref{ME1f} can be written as:
\[
y=f(\mathbf{z},\mathbf{x}(t))+g(\boldsymbol\phi)+\epsilon,
\]
where $f(\cdot)$ is the fixed-effects part and $g(\cdot)$ is the random-effects part. In this section, we will discuss the parameter estimation for both parts as well as the prediction using Model~\eqref{ME1f}.

\subsection{Model learning}
Suppose that we have selected $M$ scalar covariates and $J$ functional covariates from the candidate variables for the functional regression fixed-effects part. Let us denote the set of variables, including both functional and scalar ones, as $\tilde{\mathbf{x}}$. By ignoring the random effect part, the parameters in the fixed-effects part are estimated from \fls\ algorithm as a by product. As discussed in Section~\ref{Iter}, in the $k$-th iteration, we obtain one coefficient $\boldsymbol\beta^{(k)}$ for the covariates in the regression equation from calculating the direction vector $u^{(k)}$.  Since we used canonical correlation analysis in the calculation (Eqn~\eqref{fccaCoefGen}), this coefficient is 
\[
\boldsymbol\beta^{(k)}=\frac{P_{X,X}^{-1} V_{X,r^{(k)}}}{\rho||r^{(k)}||_2} .
\]

This coefficient is the same as the regression coefficient in this iteration up to a constant $\alpha$. Thus the regression coefficient $\tilde{\boldsymbol\beta}^{(k)}$ in the $k$-th iteration is 
\[
\tilde{\boldsymbol\beta}^{(k)}=\alpha\frac{P_{X,X}^{-1} V_{X,r^{(k)}}}{\rho||r^{(k)}||_2} \left\Vert \frac{\rho||r^{(k)}||_2}{P_{X,X}^{-1} V_{X,r^{(k)}} \tilde{\mathbf{x}}} \right\Vert_2.
\]

When the \fls\ stops at $K=J+M+1$, we can have an estimation of the the final regression coefficient for the fixed-effects:
\[
\hat{\boldsymbol\beta}=\sum_{k=1}^K \tilde{\boldsymbol\beta}^{(k)}.
\]

Once the parameters in the fixed-effects are estimated, we can estimate the hyper-parameters involved in the Gaussian process random-effects part. Note that 
\begin{align*}
r&=y-\hat{f}(\mathbf{z},\mathbf{x}(t))
=g(\boldsymbol\phi)+\epsilon;\\
g(\boldsymbol\phi)&\sim GP(0,\kappa(\boldsymbol\phi,\boldsymbol\phi'; \boldsymbol\theta)),
\end{align*}
where $\hat{f}(\mathbf{z},\mathbf{x}(t))$ is the fitted value, and $\kappa(\cdot,\cdot;\boldsymbol\theta)$ is a covariance kernel function with hyper-parameters $\boldsymbol\theta$. Taking squared exponential kernel function as an example, the covariance between sample $\boldsymbol\phi$ and $\boldsymbol\phi'$ is
\[
\kappa(\boldsymbol\phi,\boldsymbol\phi'; \boldsymbol\theta)=v_1 \exp \left( -\frac{1}{2} \sum_{h=1}^H w_h (\boldsymbol\phi-\boldsymbol\phi')^2 \right)+\delta \sigma^2,
\]
where $\delta$ is an indicator function and the hyper-parameters $\boldsymbol\theta=(v_1, w_1, \ldots, w_H, \sigma)$. 

There are many different types of covariance kernel functions. They are designed to fit in different situations; for more details, see \cite{Rasmussen:2006, Shi:2011}. We use the empirical Bayesian method to estimate the  hyper-parameters. 

The fitted value of $g(\boldsymbol\phi)$ can be calculated easily once the hyper-parameters are estimated from the data collected from all the subjects. Suppose data at total $D$ visits are recorded for a particular subject, the $D\times D$ covariance matrix of $g(\boldsymbol\phi)$ at $D$ visits is denoted by $\mathbf{c}$, each element calculated from a covariance kernel function with the estimated hyper-parameters. The fitted values are given by 
\[
\hat{\mathbf{g}}=\mathbf{c}(\mathbf{c}+\sigma^2\mathbf{I})^{-1}r,
\]
and
\[
\var(\hat{\mathbf{g}}) =\sigma^2(\mathbf{c}+\sigma^2\mathbf{I})^{-1}\mathbf{c}.
\]
More accurate estimates should be calculated by repeating the following iterative procedure until convergence.
\begin{enumerate}
\item Let $\tilde{y}=y-\hat{\mathbf{g}}=f(\mathbf{z},\mathbf{x}(t))+\epsilon$. Given the estimation of $\boldsymbol\theta$ and $\hat{\mathbf{g}}$, this is a fixed-effects scalar-on-function regression model, we can estimate all the parameters using any methods discussed in Section~\ref{integretion} and \ref{cca:scalar.vs.func}.
\item Let $r=y-\hat{f}(\mathbf{z},\mathbf{x}(t))=g(\boldsymbol\phi)+\epsilon$. Given the estimate of $\boldsymbol\beta (t)$ and $\boldsymbol\gamma$, we can update the estimation of $\boldsymbol\theta$ and calculate the fitted value of $g(\boldsymbol\phi)$ as discussed above. 
\end{enumerate}

\subsection{Prediction}
It is straightforward to calculate the prediction for the fixed-effects part. If we want to calculate the prediction of the random-effects part at a new point for a subject where we have already recorded data, i.e. forecast their future recovery level, the predictive mean and variance are given by (see Chapter 2 in \cite{Shi:2011}):
\[
\hat{y}^*=c^{*T}(\mathbf{c}+\sigma\mathbf{I})^{-1}r,~r=y-\hat{f},
\]
and
\[
\var(\hat{y}^*)=\kappa(\phi^*,\phi^*)-c^{*T}(\mathbf{c}+ \sigma^2\mathbf{I})^{-1} c +\sigma^2,
\]
where $\phi^*$ is the covariate corresponding to the new data point, $\mathbf{c}$ is the covariance matrix of $g(\boldsymbol\phi)$ calculated from the $D$ observed data pints, $c^*$ is a $D\times 1$ vector with elements $\kappa(\phi^*,\phi_d)$, i.e., the covariance of $g(\boldsymbol\phi)$ between the new data points and the observed data points.

For a new subject or patient, we can just use the prediction calculated from the fixed-effects part, and update it once we record data for the subject. An alternative way is to calculate the random-effects part using the following way: $\hat{y}^*=\sum_{i=1}^Nw_i\hat{y}_i^*$, where $\hat{y}_i^*$ is the prediction as if the new data point for the $i$-th subject, $w_i$ is the weight which takes larger values if the new subject has similar conditions to the $i$-th subject (see \cite{Shi:2008}).

\section{Analysis of 3D Kinematic Data on Arm Movement}\label{NumericalComparisonHICF}
We analyse the data collected from 70 stroke survivors in this section. The patients were allocated into the acute group if they had stroke less than one month previously and the chronic group if it occurred more than 6 months previously. This is because that the the rate of recovery at these two time points after stroke is substantially different.

\begin{figure}[!htp]
\centering
\begin{subfigure}{.5\textwidth}
  \centering
  \includegraphics[width=.8\linewidth]{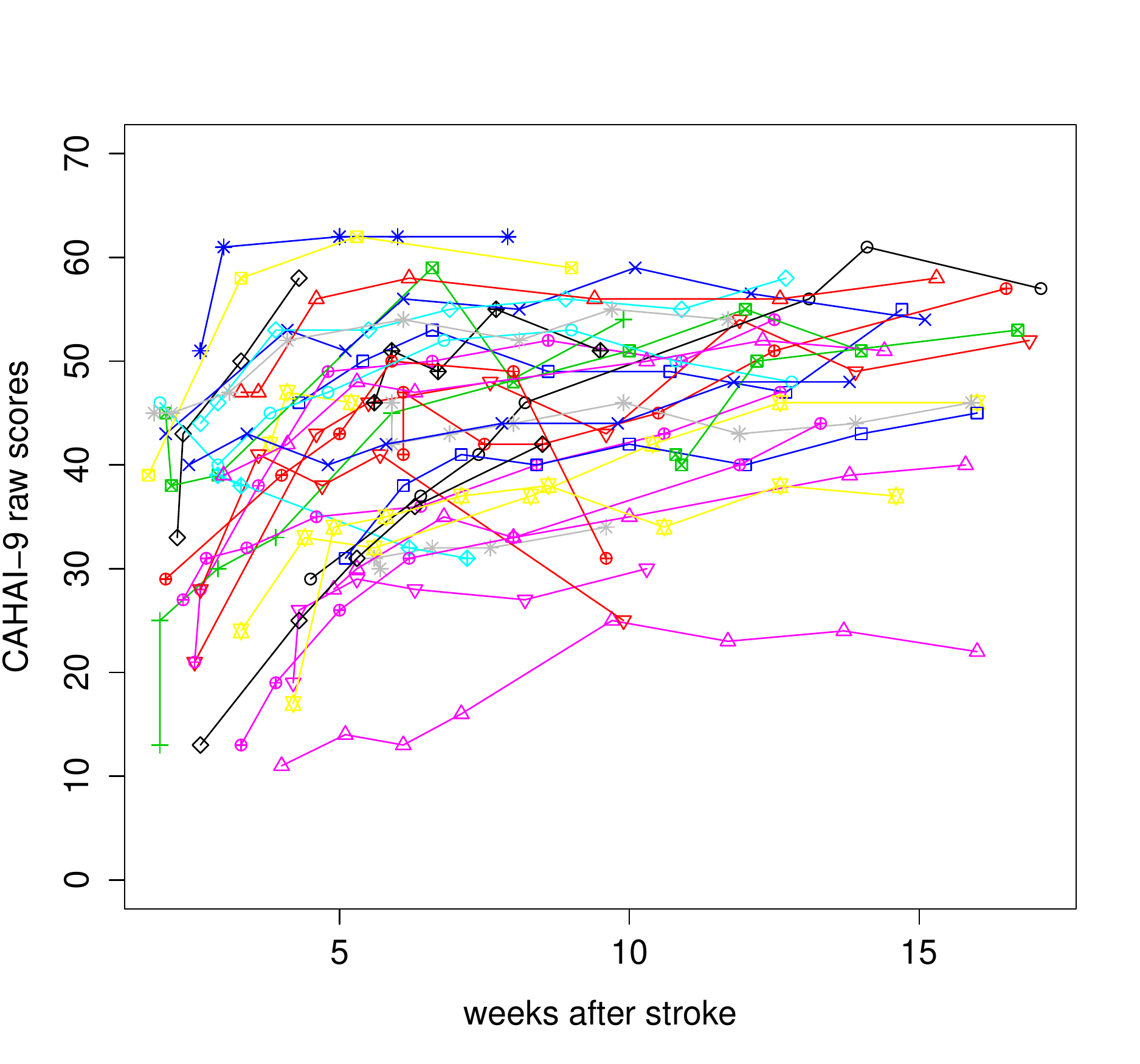}
  \caption{Acute patients} 
\end{subfigure}%
\begin{subfigure}{.5\textwidth}
  \centering
  \includegraphics[width=.8\linewidth]{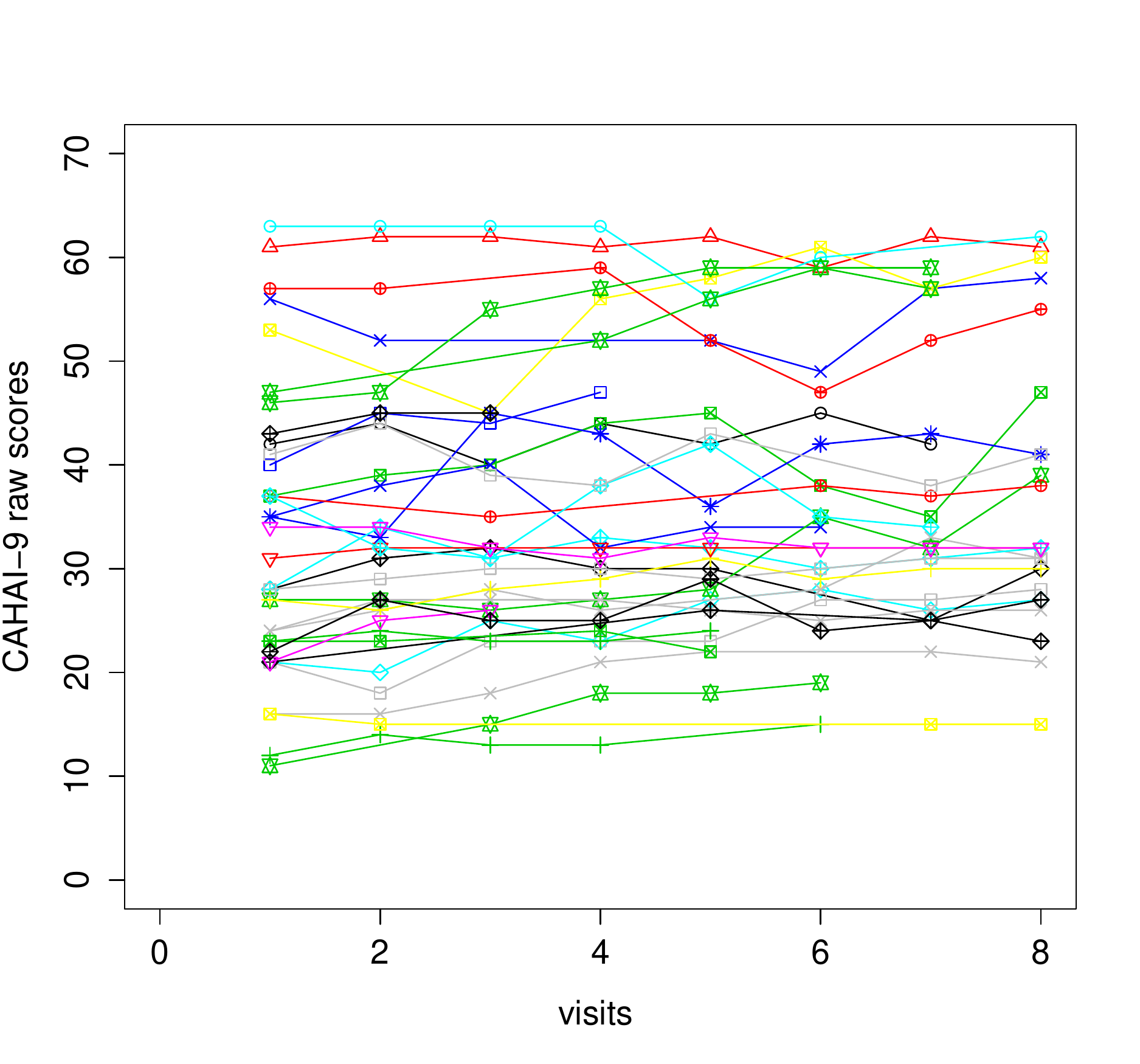}
  \caption{Chronic patients}
\end{subfigure}
\caption{Dependence level (CAHAI) against assessment time}
\label{CAHAIvsTime}
\end{figure}

Data were collected using a longitudinal design, up to eight assessments over 3 months for each patient. The first four assessments were arranged weekly, and the following fours were arranged fortnightly. For each patient, the first assessment gave the baseline dependence level, i.e. CAHAI assessment. In the following up to seven assessments, patients were asked to do the assessment game as well as the CAHAI assessment. Figure~\ref{CAHAIvsTime} shows the CAHAI values against time for acute and chronic patients respectively. Note that in the plots, the x axis for acute patients is the time since stroke, and for chronic patients is the visit time. This is because the time since stroke for chronic patients varies from a few months to a few years, which would be difficult to use in the visualization. It is clear to see that the acute patients have an increasing trend, while the trend for chronic patients is quite stable. 

A state-of-the-art prototype wireless video game controller (Sixense Entertainment: \url{http:
//www.sixense.com/}) was used by patients when they played the assessment game at home. Three dimensional position data and four dimensional orientation data (quaternion) were collected for each of the 38 movements. The samples of the raw data look very different since different patients may stand at different positions and angles with respect to the base unit, have different arm length and repeat the movement for a different number of times. Preprocessing, including calibration, standardization, segmentation, smoothing and registration were carried out before modelling. Fig~\ref{fig:la05_raw1}, \ref{fig:la05_cut_st} and \ref{fig:la05CutStSmooth} in Appendix~\ref{PrepareLA05} show the preprocessing for 10 samples from one movement. We use the data after all the preprocessing as functional variables. We also include kinematic variables in the model. They are the summary statistics of the calibrated and standardized data. For more details of the the kinematic variables, see \cite{Shi:2013}.

We have a large number of scalar and functional variables. However, both types of variables contain many missing data. They might come from the missing assessments, or the failure of doing certain movements. Due to this problem, we reduce the number of candidate variables and remove a few samples. For acute patients, there are 173 samples from 34 patients with 72 functional variables from 10 movements and 68 kinematic scalar variables from 17 movements; for chronic patients, there are 196 samples from 36 patients with 60 functional variables from 8 movements and 68 kinematic scalar variables from 19 movements. We also include the time since stroke and the baseline measurement in the candidate variables.

We apply the functional LARS algorithm to select the variables for model Eqn~\eqref{ME1f} using the whole data. The selection is done with only the linear part in the model. In other words, the variable selection is based on Model~\eqref{FSlinear}. Table~\ref{StoppingRealFFE} shows the values from our stopping rules from each of the iteration from unmodified \fls. For both types of patients, the first few variables selected by \fls\ are much more informative than the others. This is reflected by the first few distances $\alpha$ being larger than the later ones. The stopping points for the acute and chronic model could be at the ninth and eighth iterations, respectively. The stopping point for acute model is clear, but not for the chronic model. However, as the first variable is overwhelming all the others, the exact accurate stopping point is not important.

\begin{table}[!htp]
\begin{adjustwidth}{-0.5cm}{}
\begin{tabular}{r|rrrrrrrrrr}
  \hline
Acute & 2 & 3 & 4 & 5 & 6 & 7 & 8 & 9 & 10 & 11 \\ 
  \hline
$CD_{\times 1000}$ & 72.21 & 91.71 & 56.99 & 30.72 & 37.19 & 17.54 & 41.76 & 44.86 & 0.4 & 0.69 \\  
$\alpha_{\times 1000}$ & 141.36 & 213.35 & 179.92 & 88.31 & 135.65 & 71.06 & 133.26 & 176.10 & 3.48 & 2.89  \\ 
  \hline
  \hline
Chronic & 2 & 3 & 4 & 5 & 6 & 7 & 8 & 9 & 10 & 11 \\ 
  \hline
$CD_{\times 1000}$ & 327.62 & 16.33 & 11.81 & 5.73 & 10.59 & 6.67 & 8.37 & 0.02 & 10.89 & 0.7 \\ 
$\alpha_{\times 1000}$ & 830.40 & 48.27 & 47.60 & 24.90 & 55.23 & 29.73 & 44.03 & 0.21 & 57.31 & 3.23 \\ 
   \hline
\end{tabular}
\end{adjustwidth}
\caption{The changes of the stopping criteria with the iterations for \fls\ in both acute and chronic models.}
\label{StoppingRealFFE}
\end{table}
\normalsize

For acute patients, modified functional \lars\ further removes three variables from the regression equation, including two kinematic variables and one functional variable. For chronic patients, modified functional \lars\ removes one kinematic variable and two functional variables from the regression equation. In the following analysis, results from the variables selected from both unmodified and modified \fls\ will be used for comparison.

As we discussed in Section~\ref{section:inference}, we learn the model iteratively using one iteration approximation. In the step of learning fixed-effects, the parameters are recalculated using only the variables from the training data. Because the variables are all pre-selected, we estimate the coefficients by \fls\ and stop after all the variables are in the regression equation. 

To include random-effects part, several choices of variables with Gaussian process were compared, and the final model includes the time after stroke for both types of patients together with one and two kinematic variables for the acute and chronic patients, respectively.

\subsection{Numerical Comparison}
We split 34 acute patients into five folds with seven patients per fold for the first four folds, and six patients in the last fold. Similarly we split 36 chronic patients into five folds and assign eight patients into the last fold. Five-fold cross validation is then applied. More specifically, we do the predictions for one of the folds by using models learned with the other four folds. Patients were randomly selected for splitting into the folds. We repeated this process 500 times for each patient group and compared the models by using the root-mean-squared error between the prediction and the actual observations.

We detail in Table~\ref{tab:ThreeFm} the models tested. These include models of only scalar variables and those including both scalar and functional variables. Both these models were extended to include non-linear random effects using a Gaussian process, with and without modification II. The Random effects included either time of assessment only or both time of assessment and kinematic variables. The results are summarised in Table~\ref{ModelCompCAHAI}.

\begin{table}[t]
\centering
\begin{tabular}{r|r|c|c}
\hline
\vspace{-10pt} 
\multirow{3}{*}{$\mbox{FE}$}& &\multicolumn{2}{c}{ }\\
 & model & \multicolumn{2}{c}{$y=\beta_0+y^{(0)}\gamma^{(0)}+\mathbf{z}\boldsymbol\gamma+\epsilon$} \\ [1.05ex]
\cline{2-4}
\vspace{-10pt} &  &  & \\ 
 & parameters & $\boldsymbol\phi{}_{\mbox{fix}}=(1,y^{(0)},\mathbf{z})$ & - \\ [1.05ex]
\hline
\vspace{-10pt} 
\multirow{3}{*}{$\mbox{ME}_1$}& &\multicolumn{2}{c}{ }\\
 & model & \multicolumn{2}{c}{$y=\beta_0+y^{(0)}\gamma^{(0)}+\mathbf{z}\boldsymbol\gamma+g(\boldsymbol\phi)+\epsilon$} \\ [1.05ex]
\cline{2-4}
\vspace{-10pt} &  &  & \\ 
 & parameters & $\boldsymbol\phi{}_{\mbox{fix}}=(1,y^{(0)},\mathbf{z})$ & $\boldsymbol\phi=t$ \\ [1.05ex]
\hline
\vspace{-10pt}\multirow{3}{*}{$\mbox{ME}_2$} & &\multicolumn{2}{c}{ }\\
 & model & \multicolumn{2}{c}{$y=\beta_0+y^{(0)}\gamma^{(0)}+g(\boldsymbol\phi)+\epsilon$} \\ [1.05ex]
\cline{2-4}
\vspace{-10pt} &  &  & \\ 
 & parameters & $\boldsymbol\phi{}_{\mbox{fix}}=(1,y^{(0)})$ & $\boldsymbol\phi=(t,\mathbf{z})$ \\ [1.05ex]
\hline
%
\vspace{-10pt} 
\multirow{2}{*}{$\mbox{fFE}$}& &\multicolumn{2}{c}{ }\\
 & model & \multicolumn{2}{c}{$y=\beta_0+y^{(0)}\gamma^{(0)}+t\gamma^{(t)}+\mathbf{z}\boldsymbol\gamma+\sum_{j=1}^J\int x_j(t)\beta_j(t) dt+\epsilon$} \\
\cline{2-4}
\vspace{-10pt} &  &  & \\ 
 & parameters & $\boldsymbol\phi{}_{\mbox{fix}}=(1,y^{(0)},t,\mathbf{z},x_j(t))$ & - \\ [1.05ex]
\hline
\vspace{-10pt}\multirow{2}{*}{$\mbox{fME}_1$} & &\multicolumn{2}{c}{ }\\
 & model & \multicolumn{2}{c}{$y=\beta_0+y^{(0)}\gamma^{(0)}+\mathbf{z}\boldsymbol\gamma+\sum_{j=1}^J\int x_j(t)\beta_j(t) dt+g(\boldsymbol\phi)+\epsilon$} \\ [1.05ex]
\cline{2-4}
\vspace{-10pt} &  &  & \\ 
 & parameters & $\boldsymbol\phi{}_{\mbox{fix}}=(1,y^{(0)},\mathbf{z},x_j(t))$ & $\boldsymbol\phi=(t)$ \\ [1.05ex]
\hline
\vspace{-10pt} \multirow{2}{*}{$\mbox{fME}_2$}& &\multicolumn{2}{c}{ }\\
 & model & \multicolumn{2}{c}{$y=\beta_0+y^{(0)}\gamma^{(0)}+\sum_{j=1}^J\int x_j(t)\beta_j(t) dt+g(\boldsymbol\phi)+\epsilon$} \\ [1.05ex]
\cline{2-4}
\vspace{-10pt} &  &  & \\ 
 & parameters & $\boldsymbol\phi{}_{\mbox{fix}}=(1,y^{(0)})$ & $\boldsymbol\phi=(t,\mathbf{z})$ \\[1.05ex] 
\hline
\end{tabular}
\caption{The models: three mixed-effects models with only scalar variables, one fixed-effects functional regression model and three mixed-effects models with mixed scalar and functional variables. Function $g(\cdot)$ is the random-effects using Gaussian process.}
\label{tab:ThreeFm}
\end{table}

\begin{table}[t]
\begin{adjustwidth}{-.5cm}{}
\begin{tabular}{c|c|ccc|ccc|ccc}
\hline
\multicolumn{2}{c|}{} &\multicolumn{3}{c|}{Only Scalar} & \multicolumn{3}{c|}{Unmodified fLARS} & \multicolumn{3}{c}{Modified fLARS}\\
  \hline
 & & lm & $\mbox{ME}_1$ & $\mbox{ME}_2$ & $\mbox{fFE}$ & $\mbox{fME}_1$ & $\mbox{fME}_2$ & $\mbox{fFE}$ & $\mbox{fME}_1$ & $\mbox{fME}_2$ \\ 
  \hline
\hline
\multirow{2}{*}{Acute}& RMSE & 6.985 & 6.614 & 7.565 & 6.212 & 6.212 & 6.507 & 6.200 & \bf{6.061} & 6.513 \\ 
& SD & 0.229 & 0.179 & 0.141 & 0.224 & 0.220 & 0.237 & 0.186 & 0.176 & 0.198 \\ 
\hline
\multirow{2}{*}{Chronic}& RMSE & 3.904 & 4.047 & 4.270 & \bf{3.649} & 3.749 & 4.229 & 3.698 & 3.854 & 4.197 \\ 
& SD & 0.106 & 0.105 & 0.080 & 0.094 & 0.094 & 0.120 & 0.090 & 0.095 & 0.108 \\ 
   \hline 
\end{tabular}
\caption{Model comparison using prediction RMSE based on 500 replications of 5-fold cross-validation.}
\label{ModelCompCAHAI}
\end{adjustwidth}
\end{table}

For acute patients, the best model is the mixed-effects model with variables selected by modified functional LARS. It has mixed functional and scalar variables in the fixed-effects and kinematic variables in the random-effects. By including functional variables, the prediction accuracy improved significantly; by using modification II in the selection and nonlinear random-effects, the prediction accuracy improved further. For chronic patients, the best model is the fixed-effects model using unmodified functional LARS. By including the functional variables, the prediction accuracy has a clear improvement but inclusion of random-effects has no improvement.

It is clear that by including functional variables in the modelling, we can improve the prediction accuracy for both acute and chronic patients. For both types of patients, the difference between functional LARS with or without Modification II is small. The effect of including the Gaussian process random-effects depended on the data set. The chronic patients all have very small changes during the assessment period, thus almost all the variation in the data set can be captured by the fixed-effects model, while the acute patients have very different patterns, requiring the random-effects model to capture the between patients variation.

\section{Conclusion and Discussion}\label{conclusion}
In this paper, we proposed functional least angle regression, a new variable selection algorithm, for linear regression with scalar response and mixed functional and scalar covariates. This algorithm is efficient and accurate. The correlation measure used in the algorithm is from a modified functional canonical correlation analysis. It gives a correlation and a projection simultaneously. Due to the usage of the tuning parameters, conventional stopping rules fail in this algorithm. We proposed two new stopping rules. The simulation studies and the real data analysis show that the performance of this new algorithm together with the new stopping rules performs well in complex data. The integrations involved in the calculation for functional variables are carried out by three different ways: the representative data points, the Gaussian Quadrature and the basis functions. The second method is new and it turns out to be as accurate as other two methods and it is efficient in the calculation. Further research is justified to define the optimal representative data points for functional variables.

This algorithm, however, has the potential for further improvement. As we explained in Section~\ref{cca:scalar.vs.func}, functional canonical correlation analysis is just one choice of the many correlation measures we could use, for example non-linear canonical correlation \citep{van1994overals} and kernel canonical correlation \citep{lai2000kernel}. The new stopping rules we proposed are based on the logic of the algorithm, while the conventional stopping rules are normally based on information criteria. Additional research could investigate if the stopping rules can be enhanced by including a measure of the information in the selected variables. The nonlinear random-effects, based on a nonparametric Bayesian approach with a Gaussian process prior, has been shown to be a flexible method for a complex model of heterogenous subjects, that can cope with multi-dimensional covariates.

An R package, named as fLARS, has been developed and will be available in R CRAN very soon. 

\section*{Acknowledgements}
This publication presents independent research commissioned by the Health Innovation Challenge Fund (Grant number HICF 1010 020), a parallel funding partnership between the Wellcome Trust and the Department of Health. The views expressed in this publication are those of the authors and not necessarily those of the Wellcome Trust or the Department of Health.

\addcontentsline{toc}{section}{Appendices}
\section{Appendix}
\subsection{Use Gaussian quadrature to approximate the integration}\label{GQapp}
The integration Eqn~\eqref{Eqn:IntIllu} can be written as:
\begin{align*}
u&=\int x(t)\beta(t) dt=\int f(t) dt\\
&\approx \sum_{i=1}^nw_if(t_i)=\sum_{i=1}^nw_ix(t_i)\beta(t_i)
\end{align*}

Similarly, the integration related to the penalty function can be written as:
\begin{align*}
\int [\beta''(t)] dt&=\int f(t) dt\approx \sum_{i=1}^nw_if(t_i)\\
&=\sum_{i=1}^nw_i\beta''(t_i)\beta''(t_i)=\sum_{i=1}^nw_i\beta''(t_i)\beta''(t_i)
\end{align*}

We also define the second order derivative of the functional coefficient by $L\mytilde{\beta}^T$. Since the discrete data points in Gaussian quadrature are unevenly spaced, we need a general solution of finite difference. Suppose $f(t)$ is the original function, which has observations on $t=t_j$, where $j=1,\ldots, J$. By Taylor expansions, 
\begin{align*}
f(t_{j+1})=f(t_j)+(t_{j+1}-t_j)f'(t_j)+\frac{(t_{j+1}-t_j)^2}{2!}f''(t_j)\ldots\\
f(t_{j-1})=f(t_j)-(t_j-t_{j-1})f'(t_j)+\frac{(t_j-t_{j-1})^2}{2!}f''(t_j)\ldots.
\end{align*}

Thus the centred differences formula for second order derivative is:
\begin{align*}
f''(t_j)=\frac{f(t_{j+1})(t_j-t_{j-1})-f(t_j)(t_{j+1}-t_{j-1})+f(t_{j-1})(t_{j+1}-t_j)}{(t_{j+1}-t_j)(t_{j}-t_{j-1})\frac{t_{j+1}-t_{j-1}}{2}}.
\end{align*}

This gives the weight for the entries of the matrix $L$:

\[
\hspace{-.2in}
L=\left(\begin{array}{cccccc}
\frac{2}{(t_3-t_2)(t_3-t_1)} &  -\frac{2}{(t_3-t_2)(t_2-t_1)} & \frac{2}{(t_2-t_1)(t_3-t_1)} & 0  & 0 & \ldots\\
0  & \frac{2}{(t_4-t_3)(t_4-t_2)} &  -\frac{2}{(t_4-t_3)(t_3-t_2)} & \frac{2}{(t_3-t_2)(t_4-t_2)} & 0 & \ldots\\
0 & 0  & \ddots &  \ddots & \ddots & \ldots\\
 0 & 0 & 0 & \ldots & \ldots & \ldots 
\end{array}\right),
\]

\subsection{Calculation of the degrees of freedom for fLARS}\label{dfFLARS}
Recall Eqn~\eqref{NewRes}, the residual after iteration $k$ can be written as
\begin{align*}
r^{(k+1)}=r^{(k)}-\alpha_k u^{(k)}
\end{align*}
where $u$ is the direction vector, calculated by:
\begin{align*}
u^{(k)}=\frac{H_k r^{(k)}}{{\sd}(H_k r^{(k)})}.
\end{align*}
Therefore, the true ``hat'' matrix at iteration $k$ is :
\begin{align*}
H_k^*=\frac{H_k \alpha_k}{\text{\sd}(H_k r^{(k)})},
\end{align*}
where $H_k$ is from Eqn~\eqref{fccaCorGen} and Eqn~\eqref{fccaCoefGen}. More specifically, if $\mathbf{X}$ is the matrix which combines all selected variables, $H_k=\mathbf{X}W (W^T \mathbf{X}^T\mathbf{X}W+ \lambda_1 W_2 +\lambda_2 W^TW)^{-1} W^T\mathbf{X}^T$. The residual after iteration $k$ becomes:
\begin{align*}
r^{(k+1)}=(I-H_k^*)r^{(k)}.
\end{align*}
Recursively, the fitted value after iteration $K$ with respect to the response variable is:
\begin{align*}
\hat{y}=y- r^{(K+1)}
 = y-[\prod_{k=1}^{K}(I-H_k^*)]y
 = [I-\prod_{k=1}^{K}(I-H_k^*)]y,
\end{align*}
hence the `hat' matrix $\bar{H}_K$ after iteration $K$ is 
\begin{align*}
\bar{H}_K=I-\prod_{k=1}^{K}(I-H_k^*) .
\end{align*}

We then define the degrees of freedom for functional \lars\ as follows:
\begin{align*}
\text{df}^*=\tr\Bigg(\frac{\cov(\hat{\boldsymbol\mu^*}^T,\mathbf{y}^{*T})}{\sigma^2}\Bigg)
=\tr\Bigg(\frac{\cov(\mathbf{y}^{*T}\bar{H}_k^T,\mathbf{y}^{*T})}{\sigma^2}\Bigg)
=\tr\Bigg(\frac{\bar{H}_k\mathbf{y}^*\mathbf{y}^{*T}/(n-1)}{\sigma^2}\Bigg)
\end{align*}
where $\mathbf{y}^*\mathbf{y}^{*T}/(n-1)$ is an $n\times n$ matrix. The $i, j$-th element is $\cov(y_i,y_j)$. Its value is $\sigma^2$ if $i=j$ and 0 otherwise. Hence:
\begin{align*}
\text{df}^*=\tr\Bigg(\frac{\bar{H}_k\mathbf{y}^*\mathbf{y}^{*T}/(n-1)}{\sigma^2}\Bigg) 
=\tr\Bigg(\frac{\bar{H}_k\sigma^2I}{\sigma^2}\Bigg) 
=\tr(\bar{H}_k). 
\end{align*}

\subsection{Plots of the preprocessing of one movement}\label{PrepareLA05}
\begin{figure}[!b]
\includegraphics[height=3in,width=5.6in]{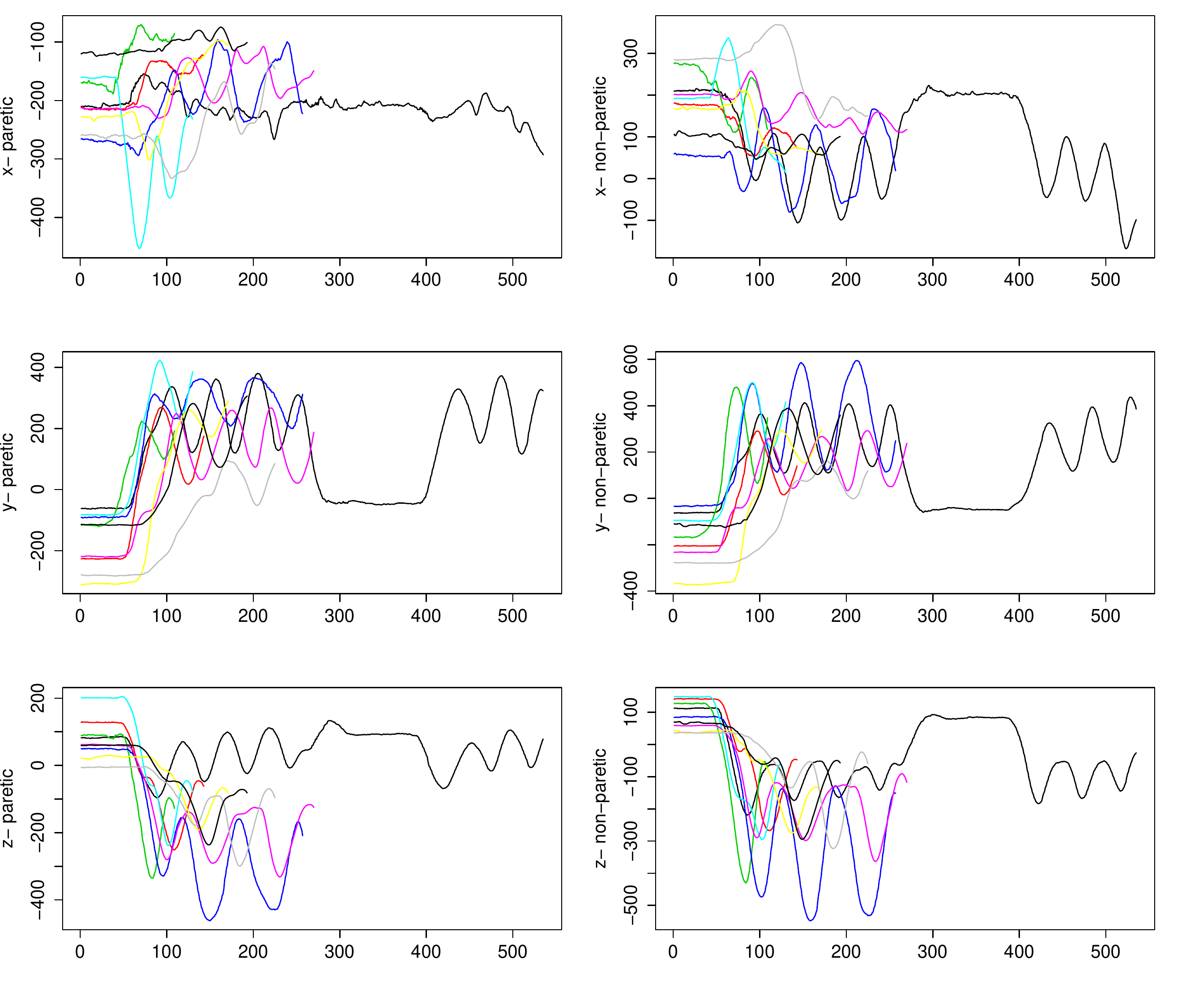}
\caption{10 samples of the raw data from the movement of Forward roll}
\label{fig:la05_raw1}
\end{figure}

\begin{figure}[t]
\includegraphics[height=3in,width=5.6in]{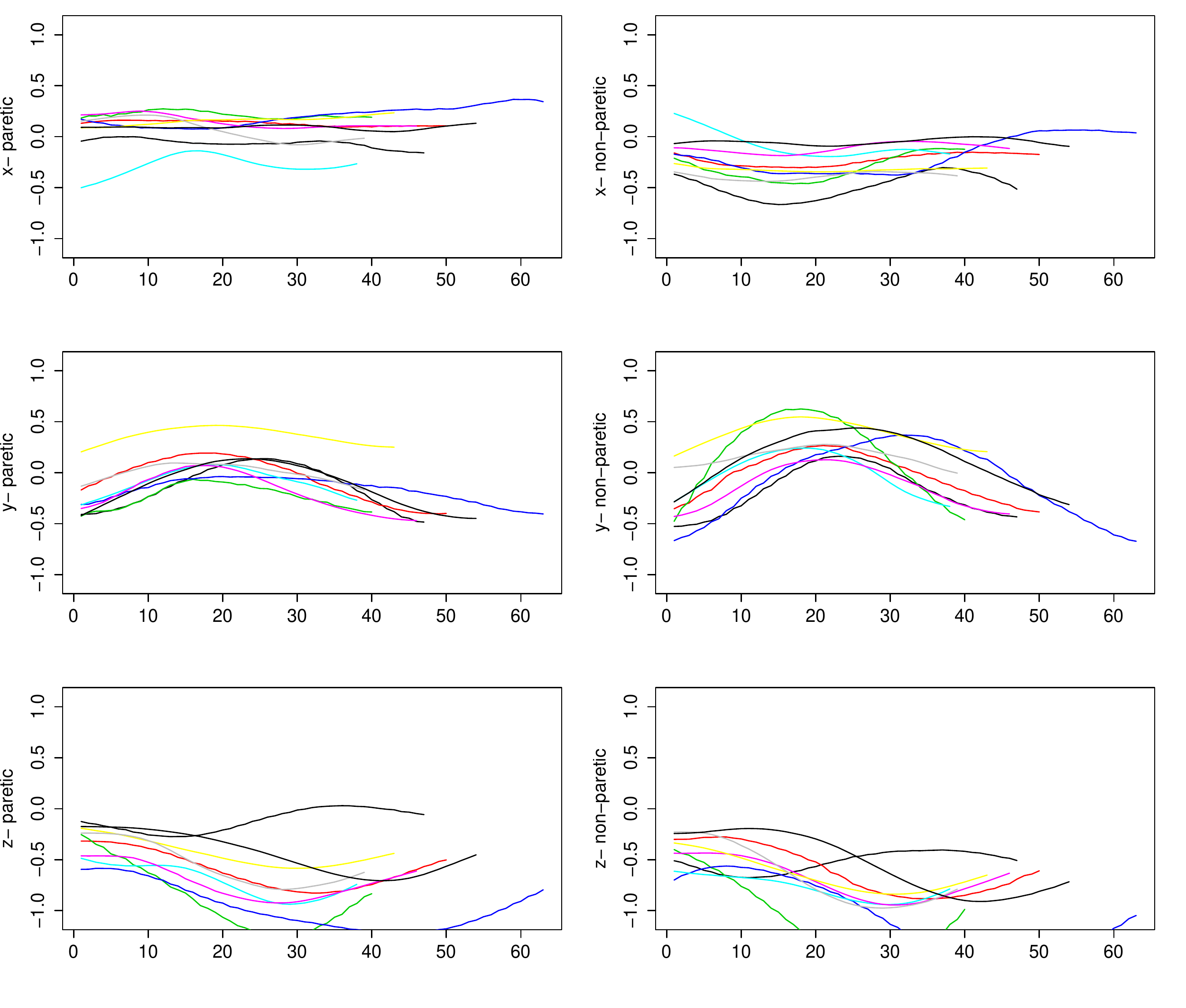}
\caption{10 samples of the standardized and segmented data from movement Forward roll}
\label{fig:la05_cut_st}
\end{figure}

\begin{figure}[b]
\includegraphics[height=3in,width=5.6in]{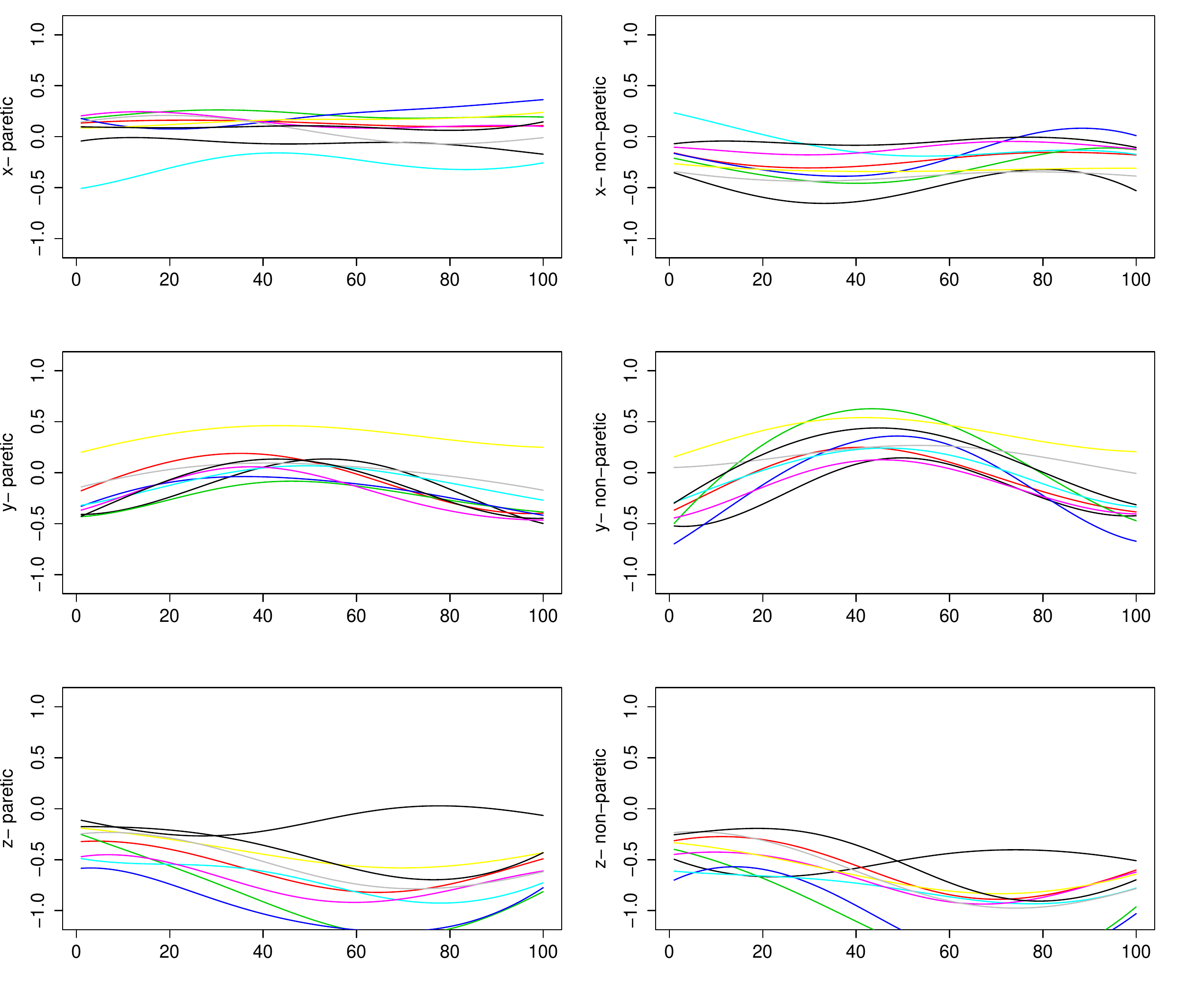}
\caption{10 samples of the data after preprocessing from movement Forward roll}
\label{fig:la05CutStSmooth}
\end{figure}

\clearpage

\bibliography{reference} 
  \bibliographystyle{plainnat}
  
\end{document}